\documentclass[%
reprint,
superscriptaddress,
amsmath,amssymb,
aps,
pra,
]{revtex4-2}
\usepackage[utf8]{inputenc}
\usepackage[english]{babel}
\usepackage{blindtext}
\usepackage{bbm}
\usepackage{graphicx}   
\graphicspath{{figs/}}   
\usepackage{dcolumn}
\usepackage{bm}
\usepackage[hidelinks]{hyperref}
\usepackage{svg}
\usepackage[section]{placeins}
\usepackage{epstopdf}
\usepackage{braket}
\usepackage{upgreek}
\usepackage{amsmath}
\usepackage{siunitx}
\usepackage{changes}
\usepackage{natbib}
\usepackage[title]{appendix}

\usepackage{multirow}
\usepackage{import}
\usepackage{calc}
\usepackage{xcolor}

\newcommand{\secref}[1]{\hyperref[#1]{{Sec.~\ref{#1}}}}
\newcommand{\chapref}[1]{\hyperref[#1]{{Ch.~\ref{#1}}}}
\newcommand{\appref}[1]{\hyperref[#1]{{App.~\ref{#1}}}}
\newcommand{\figref}[1]{\hyperref[#1]{{Fig.~\ref*{#1}}}}
\newcommand{\tabref}[1]{\hyperref[#1]{{Table~\ref*{#1}}}}

\renewcommand{\eqref}[1]{\hyperref[#1]{{Eq.~\ref*{#1}}}}

\newcommand{\revisex}[1]{{}}

\definecolor{myBrown}{HTML}{947139}
\definecolor{myDarkgreen}{HTML}{6E692A}
\definecolor{myGreen}{HTML}{00A76D}
\definecolor{myLightgreen}{HTML}{A6CE42}
\definecolor{myYellow}{HTML}{FFF200}
\definecolor{myRed}{HTML}{ED1C1A}
\definecolor{myOrange}{HTML}{F7931D}
\definecolor{myLightblue}{HTML}{00C0F3}
\definecolor{myPink}{HTML}{F6979F}
\definecolor{myBlue}{HTML}{0071BC}
\definecolor{myGold}{HTML}{FFCB04}
\definecolor{myPurple}{HTML}{A066AA}
\definecolor{myDarkgrey}{HTML}{6D6E70}
\definecolor{myLightgrey}{HTML}{9D9FA1}

\begin{document}

\title{Parametric two-qubit gates via Landau-Zener interference} 

\author{Simon~Geisert}
\affiliation{IQMT,~Karlsruhe~Institute~of~Technology,~76131~Karlsruhe,~Germany}

\author{Albert~Hertel}
\affiliation{Qruise GmbH,~66113~Saarbruecken,~Germany}

\author{Soeren~Ihssen}
\affiliation{IQMT,~Karlsruhe~Institute~of~Technology,~76131~Karlsruhe,~Germany}

\author{Zhongyi~Jiang}
\affiliation{PGI-12,~Forschungszentrum~Juelich,~52428~Juelich,~Germany}

\author{Paul~Kugler}
\affiliation{IQMT,~Karlsruhe~Institute~of~Technology,~76131~Karlsruhe,~Germany}

\author{Nicolas~Zapata}
\affiliation{IQMT,~Karlsruhe~Institute~of~Technology,~76131~Karlsruhe,~Germany}

\author{Nicolas~Gosling}
\affiliation{IQMT,~Karlsruhe~Institute~of~Technology,~76131~Karlsruhe,~Germany}

\author{Ameya~Nambisan}
\affiliation{IQMT,~Karlsruhe~Institute~of~Technology,~76131~Karlsruhe,~Germany}

\author{Yuan~Gao}
\affiliation{Institute~for~Functional~Quantum~Systems~(PGI-13),~Forschungszentrum~Jülich,~52425~Jülich,~Germany}
\affiliation{Department~of~Physics,~RWTH~Aachen~University,~52074~Aachen,~Germany}

\author{Asier~Galicia}
\affiliation{Institute~for~Functional~Quantum~Systems~(PGI-13),~Forschungszentrum~Jülich,~52425~Jülich,~Germany}
\affiliation{Department~of~Physics,~RWTH~Aachen~University,~52074~Aachen,~Germany}

\author{Jéferson~R.~Guimarães}
\affiliation{Institute~for~Functional~Quantum~Systems~(PGI-13),~Forschungszentrum~Jülich,~52425~Jülich,~Germany}
\affiliation{Department~of~Physics,~RWTH~Aachen~University,~52074~Aachen,~Germany}

\author{Yorgo~Haddad}
\affiliation{Institute~for~Functional~Quantum~Systems~(PGI-13),~Forschungszentrum~Jülich,~52425~Jülich,~Germany}
\affiliation{Department~of~Physics,~RWTH~Aachen~University,~52074~Aachen,~Germany}

\author{Marc~Neis}
\affiliation{Institute~for~Functional~Quantum~Systems~(PGI-13),~Forschungszentrum~Jülich,~52425~Jülich,~Germany}
\affiliation{Department~of~Physics,~RWTH~Aachen~University,~52074~Aachen,~Germany}

\author{Harsh~Bhardwaj}
\affiliation{Institute~for~Functional~Quantum~Systems~(PGI-13),~Forschungszentrum~Jülich,~52425~Jülich,~Germany}
\affiliation{Department~of~Physics,~RWTH~Aachen~University,~52074~Aachen,~Germany}

\author{Dmitriy~A.~Volkov}
\affiliation{Institute~for~Functional~Quantum~Systems~(PGI-13),~Forschungszentrum~Jülich,~52425~Jülich,~Germany}
\affiliation{Department~of~Physics,~RWTH~Aachen~University,~52074~Aachen,~Germany}

\author{Juan~Cereijo}
\affiliation{Institute~for~Functional~Quantum~Systems~(PGI-13),~Forschungszentrum~Jülich,~52425~Jülich,~Germany}
\affiliation{Department~of~Physics,~RWTH~Aachen~University,~52074~Aachen,~Germany}

\author{Marcello~Guardascione}
\affiliation{Institute~for~Functional~Quantum~Systems~(PGI-13),~Forschungszentrum~Jülich,~52425~Jülich,~Germany}
\affiliation{Department~of~Physics,~RWTH~Aachen~University,~52074~Aachen,~Germany}

\author{Yebin~Liu}
\affiliation{Institute~for~Functional~Quantum~Systems~(PGI-13),~Forschungszentrum~Jülich,~52425~Jülich,~Germany}

\author{Markus~Jerger}
\affiliation{Institute~for~Functional~Quantum~Systems~(PGI-13),~Forschungszentrum~Jülich,~52425~Jülich,~Germany}

\author{Pavel~Bushev}
\affiliation{Institute~for~Functional~Quantum~Systems~(PGI-13),~Forschungszentrum~Jülich,~52425~Jülich,~Germany}

\author{Frank~Wilhelm-Mauch}
\affiliation{Qruise GmbH,~66113~Saarbruecken,~Germany}
\affiliation{PGI-12,~Forschungszentrum~Juelich,~52428~Juelich,~Germany}

\author{Wolfgang~Wernsdorfer}
\affiliation{IQMT,~Karlsruhe~Institute~of~Technology,~76131~Karlsruhe,~Germany}
\affiliation{PHI,~Karlsruhe~Institute~of~Technology,~76131~Karlsruhe,~Germany}

\author{Shai~Machnes}
\affiliation{Qruise GmbH,~66113~Saarbruecken,~Germany}

\author{Mohammad~Ansari}
\affiliation{PGI-12,~Forschungszentrum~Juelich,~52428~Juelich,~Germany}

\author{Rami~Barends}
\affiliation{Institute~for~Functional~Quantum~Systems~(PGI-13),~Forschungszentrum~Jülich,~52425~Jülich,~Germany}

\author{Ioan~M.~Pop}
\email{ioan.pop@kit.edu}
\affiliation{IQMT,~Karlsruhe~Institute~of~Technology,~76131~Karlsruhe,~Germany}
\affiliation{PHI,~Karlsruhe~Institute~of~Technology,~76131~Karlsruhe,~Germany}
\affiliation{Physics~Institute~1,~Stuttgart~University,~70569~Stuttgart,~Germany}

\date{\today}

\begin{abstract}

   We propose and demonstrate gates between two superconducting qubits based on quantum interference of consecutive Landau-Zener (LZ) transitions. This gate mechanism bridges between baseband and parametric two-qubit control, enabling in situ tuning of the control frequency across a continuous interval up to hundreds of MHz. Another advantage compared to dispersive couplers is that the speed of the LZ gate is on the order of the full coupling strength. We experimentally demonstrate the gate on two platforms, a modular chiplet architecture of coupled generalized flux qubits~\cite{Ihssen2025Mar, Geisert2024Jul}, and on a monolithic transmon architecture~\cite{QSolid}.  The combination of tunability and gate speed establishes the LZ gate as a unique tool for multiplexing control pulses and interconnecting superconducting chiplet architectures.
\end{abstract}

\maketitle

\section{Introduction}


The prospect of fault-tolerant quantum computing has driven the development of coherent superconducting quantum hardware for more than a quarter century~\cite{Nakamura1999Apr,Vion2002May}, culminating in quantum processors with up to hundreds of interconnected qubits~\cite{Kim2023Jun, IBMQuantumRoadmap}.
Such systems rely on two-qubit gates which must operate orders of magnitude faster than the qubit coherence times in order to reach the fidelity threshold for error correction codes~\cite{Martinis2015Oct, Barends2014Apr}.
In circuit quantum electrodynamics~\cite{Blais2021May}, the strong electric- and magnetic-field-mediated interactions make it relatively straightforward to achieve the required coupling rates.
However, this raises the challenge of avoiding always-on interactions~\cite{Ku2020Nov, Xu2021Jun} and spurious couplings to neighbouring qubits.
The most common strategy consists in replacing static electromagnetic coupling~\cite{Pashkin2003Feb} with a new mode which mediates dispersive interactions.
If this is a fixed-frequency mode, gates can be induced via microwave-activated transitions~\cite{Majer2007Sep, Mollenhauer2025Jul, Wei2022Aug, Kurpiers2018Jun}. 
In case of a flux-tunable mode~\cite{vanderPloeg2007Feb, Bialczak2011Feb, Chen2014Nov, Quantum2020Sep}, gates can be activated via either baseband~\cite{Sung2021Jun} or parametric flux modulation~\cite{Niskanen2007May, Zhang2024May}.

In order to suppress state leakage, the coupler typically operates in the dispersive regime, which limits gate speeds to well below the vacuum Rabi rate. 
Moreover, parametric modulation of the gates must be performed within a narrow transition-frequency window set by the qubit-coupler level structure.
Here, we present a complementary approach beyond the dispersive regime~\cite{Ansari2019Jul} that enables two-qubit gates by driving at frequencies from baseband to hundreds of MHz while exploiting the full coupling strength. 
We couple two data qubits with equal strength to a coupler, which is itself a flux-tunable qubit, and we periodically modulate the coupler such that it repeatedly crosses the qubit frequencies to induce Landau–Zener (LZ) interference. 
Two-qubit gates are realized by tuning the coupler detuning, the drive amplitude and its frequency to achieve constructive interference between LZ transitions. 
Because, for any drive frequency, a corresponding set of drive amplitude and coupler detuning satisfies the LZ resonance condition, the gate can be activated at arbitrary drive frequencies.

LZ transitions have a long history in the context of quantum control: Indeed, the theoretical framework of LZ transitions and LZ interference dates back to the 1930s~\cite{Zener1932Sep, Stueckelberg1932} and, as discussed in detail in the review by Shevchenko, Ashhab and Nori~\cite{Shevchenko2010Jul}, has since been theoretically investigated and experimentally demonstrated on various platforms~\cite{Ota2018Apr, Peyruchat2025Apr, Gorelik1998Sep, Wernsdorfer2000May}. Particularly relevant to our work are recent theoretical proposals on single- and two-qubit gates in superconducting devices~\cite{Ferreyra2026Jul, Reparaz2025Aug, Ryzhov2024Sep, Caceres2023Nov}, experimental demonstrations of flux qubit~\cite{Oliver2005Dec, Oliver2009Jun} and charge qubit~\cite{Sillanpaa2006May, Silveri2015, Bjorkman2025Feb} control and, most importantly, two-qubit gates between transmon qubits~\cite{Campbell2020Dec}. 

Our goal is to expand this concept to two-qubit control including tunable couplers. In particular, we address the situation in which the tunability of the coupler exceeds the coupling strength of the qubits. In this regime, the system is well-described by the adiabatic impulse (or transfer-matrix) model~\cite{Damski2006Jun,Ashhab2007Jun}. An intuitive picture of the gate mechanisms is given by the analogy to an array of beam splitters and mirrors in a Mach-Zehnder interferometer~\cite{Oliver2005Dec, Zehnder1891}.
In this analogy, the combined choice of length, frequency and amplitude of a flux pulse applied to the coupler determines the equivalent number of beam splitters, optical path lengths, and beam splitter reflectivity. 

The manuscript has the following structure: We first introduce the theoretical concept using the adiabatic impulse model in \secref{sec:model}, where we outline the analogy to the Mach-Zehnder interferometer. In \secref{sec:experiments} we show the experimental results on two different superconducting platforms. We also discuss the potential and the limitations of Landau-Zener interferometric gate mechanism. In the appendices we provide further details on the gate derivation, error sources, and a complementary interpretation of the gate mechanism in terms of Floquet sidebands. 

\section{Theory}
\label{sec:model}

\subsection{Circuit model and level structure}

\begin{figure}[ht]
    \includegraphics[width=1.0\columnwidth]{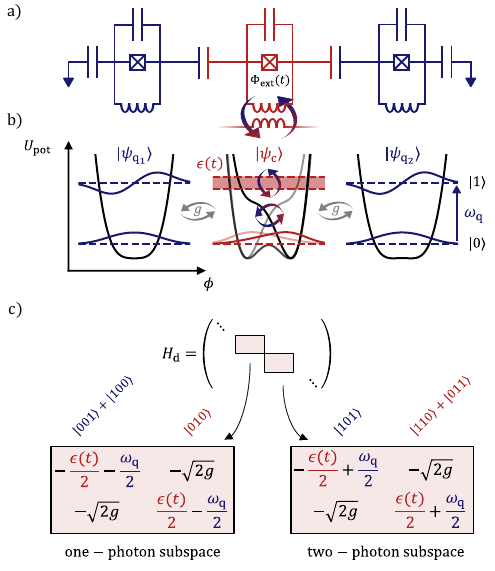}
    \caption{
\textbf{Coupled qubit array}. \textbf{a)} Array of three capacitively coupled generalized flux qubits (\cite{Yan2020Jun, Geisert2024Jul, Ihssen2025Mar}) with outermost data qubits (blue) and fast-flux tunable coupler (red) in the center. \textbf{b)} Potential energy versus superconducting phase differences $\phi$ across the junctions. The external magnetic flux threading the qubit loops is biased to $\Phi_\text{ext} = \Phi_0 / 2$, resulting in even and odd wavefunctions for the ground and excited states separated at frequency $\omega_{\mathrm{q}}$. The coupler wavefunction and detuning $\epsilon(t) = \omega_{\mathrm{c}}(t) - \omega_{\mathrm{q}}$ adiabatically follows the flux modulation $\Phi_{\mathrm{ext}}(t)$. \textbf{c)} The Hamiltonian $H_\text{d}$ contains two blocks (light red) that each describe independent avoided level crossings between the bright states (red and blue) in the one- and two-photon subspace (left and right, respectively).   } 
    \label{fig:fig1}
\end{figure}

The circuit consists of an array of three nearest-neighbour coupled superconducting qubits with equal coupling constants $g$ as visualized in \figref{fig:fig1}a. Due to their design versatility and well-balanced ratio of anharmonicity and frequency~\cite{Yan2020Jun}, we choose to implement generalized flux qubits~\cite{Geisert2024Jul} close to the quarton regime. The outer qubits serve as data qubits ($\mathrm{q_1}$, $\mathrm{q_2}$) and are tuned on resonance $\omega_{\mathrm{q}}$ via the external magnetic field, whereas the center qubit acts as the tunable coupler qubit (c) whose resonance frequency can be controlled via a fast-flux bias line by the amount $\epsilon(t)$ around $\omega_\text{q}$, see \figref{fig:fig1}b. 

The circuit layout follows the modular chiplet architecture presented in Ref.~\cite{Ihssen2025Mar}. To model the gate mechanism, we truncate the Hilbert space to the computational subspace $\ket{\psi_{\mathrm{q_1}}\psi_{\mathrm{c}}\psi_{\mathrm{q_2}}} \in \{\ket{000},..., \ket{111}\}$, such that the Hamiltonian in the local basis may be written as

\begin{equation}
\begin{aligned}
    \frac{H_{\mathrm{local}}(t)}{\hbar} = &-\frac{\omega_{\mathrm{q}}}{2} \left( \sigma_z^{\mathrm{q_1}} + \sigma_z^{\mathrm{c}} + \sigma_z^{\mathrm{q_2}} \right) - \frac{\epsilon (t)}{2} \sigma_z^{\mathrm{c}} \\
    & - g \left( \sigma_x^{\mathrm{q_1}}\sigma_x^{\mathrm{c}} + \sigma_x^{\mathrm{c}}\sigma_x^{\mathrm{q_2}} \right).
    \label{eq:H_local}
\end{aligned}
\end{equation}
While this simplified model Hamiltonian does not include the full physical level structure of the superconducting circuit, it is sufficient to explain and validate the interferometric gate mechanism.

For finite $g$ and detuned coupler $\frac{|\epsilon|}{g} \gg 1$, the eigenstates can be readily inferred from the symmetry of the system. Since the data qubits are on resonance, the computational basis of the three qubits can be rotated via the transformation matrix $S$ to form two pairs of nearly degenerate states, corresponding to the single photon ($\ket{100} + \ket{001}$ and $\ket{100} - \ket{001}$) and two-photon subspace ($\ket{110} + \ket{011}$ and $\ket{110} - \ket{011}$), see \appref{sec:derivation}. We consider the regime $g \ll \omega_{\mathrm{q}}$, ensuring that there are no transitions between different excitation number subspaces. In the antisymmetric combination the interaction with the coupler cancels, therefore we refer to these states as ``dark". The ``bright" states interact with the coupler forming the diabatic Hamiltonian $H_{\mathrm{d}} = S^T H_{\mathrm{local}} S$, which is block-diagonal, as shown in \figref{fig:fig1}c.

\begin{figure}[htbp]
    \includegraphics[width=1.0\columnwidth]{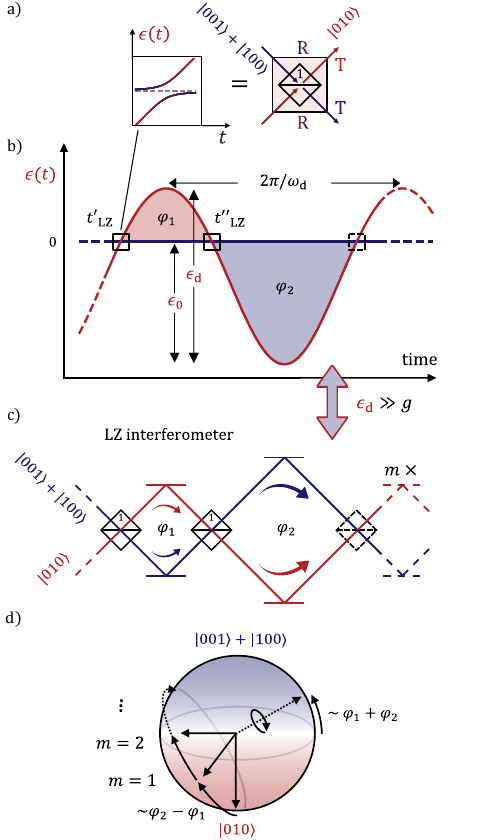}
    \caption{
\textbf{Landau-Zener interferometer}. \textbf{a)} Sweeping the coupler frequency (red) through the data qubit frequency (blue) yields LZ transitions between the bright states which are modeled as beam splitters with reflection and transmission coefficients $R$ and $T$. The antisymmetric dark state is decoupled and depicted as a gray dashed line. \textbf{b)} Upon periodic modulation of the coupler frequency with period $2\pi / \omega_{\mathrm{d}}$, strength $\epsilon_{\mathrm{d}}$ and offset $\epsilon_{\mathrm{0}}$, consecutive LZ transitions and time intervals of free phase evolution can be identified. The LZ transitions are highlighted as black rectangles. The red and blue shaded areas highlight the acquired relative phases $\varphi_1$ and $\varphi_2$. \textbf{c)} In analogy to the Mach-Zehnder interferometer, the sequence of LZ transitions is modeled as a series of $m$ interferometers. \textbf{d)} The state evolution is illustrated on the Bloch sphere of the effective two-level system consisting of the bright state and the coupler. The sum and difference of the relative phases between LZ transitions determine the rotation axis and angle. The passage of $m$ interferometers rotates the Bloch vector $m$ times.}
    \label{fig:fig2}
\end{figure}

\subsection{Landau-Zener interferometry}

When the coupler is tuned linearly through the bright states, see \figref{fig:fig2}a, the transition is well-described by a LZ transfer matrix which resembles a beam splitter interaction~\cite{Shevchenko2010Jul}
\begin{equation}
N = 
\begin{pmatrix}
T & R e^{i\varphi_s} \\
-R e^{-i\varphi_s} & T
\end{pmatrix},
\label{eq:beam_splitter}
\end{equation}
where $T = \sqrt{P}$ and $R = \sqrt{1-P}$ are the transmission and reflection amplitudes determined by the Landau-Zener formula
\begin{equation}
    P = e^{-2\pi \delta} \qquad \text{with} \qquad \delta = \frac{\Delta^2}{4v}.
    \label{eq:LZ_probability}
\end{equation}
The avoided crossing gap is $\Delta = 2\times\sqrt{2}g$ (see \appref{sec:derivation}). The tunneling probability $P$ decreases with increasing adiabaticity $\delta$, which is characterized by the velocity $v$ with which the crossing is traversed. In our case, the velocity is given by how fast the coupler is modulated through the crossing, i.e. 
\begin{equation}
    v = \left. \frac{\partial \epsilon}{\partial t} \right|_{t=t_{\mathrm{LZ}}}
    \label{eq:lz_velocity}
\end{equation}
evaluated at the time of the crossing $t_{\mathrm{LZ}}$. The Stokes phase acquired upon reflection weakly depends on $\delta$ and in case of $\delta \ll 1$ is close to $\varphi_s \approx \frac{\pi}{4}$. To determine the LZ parameters in \eqref{eq:LZ_probability}, the exact shape of the coupler frequency modulation should be taken into account, as detailed in \appref{sec:nonlinear}.

We implement Landau-Zener interferometry by periodically modulating the coupler with a frequency $\omega_{\mathrm{d}}~=~2\pi/T$ and sufficiently large amplitude $\epsilon_\text{d}$ such that it crosses the bright states, as shown in \figref{fig:fig2}b. Within one period $T$ there are two crossing points, $t'_{\mathrm{LZ}}$ and $t''_{\mathrm{LZ}}$, at which the system undergoes subsequent LZ transitions, $N$ and $N^T$, in opposite directions. In between transitions, the system evolves adiabatically according to
\begin{equation}
U_{i} = 
\begin{pmatrix}
e^{i\frac{\varphi_{i}}{2}} & 0 \\
0 & e^{-i\frac{\varphi_{i}}{2}}
\end{pmatrix},
\label{eq:unitary_evolution}
\end{equation}
with 
\begin{equation}
    \varphi_1 = \int_{t'_{\mathrm{LZ}}}^{t''_{\mathrm{LZ}}} \epsilon (t)\text{d}t 
\quad \mathrm{and} \quad \varphi_2 = \int_{t''_{\mathrm{LZ}}}^{t'_{\mathrm{LZ}}+T} \epsilon (t)\text{d}t ,
\label{eq:phase_differences}
\end{equation} as indicated by the colored areas in \figref{fig:fig2}{b}. Therefore, the LZ interference matrix after $m$ periods is described by the so-called adiabatic impulse model~\cite{Damski2006Jun, Ashhab2007Jun, Shevchenko2010Jul, Oliver2005Dec}
\begin{equation}
     M^m = \left(U_{2}N^TU_{1}N \right)^{m}.
    \label{eq:lzsm_cycle}
\end{equation}

\autoref{eq:lzsm_cycle} allows the evolution over $m$ drive cycles to be interpreted as a time-domain sequence of Mach-Zehnder interferometers~\cite{Zehnder1891}. As schematized in \figref{fig:fig2}c, the sequence consists of a total of $2m$ beam splitters with relative phases $\varphi_1$ and $\varphi_2$ between the two arms (cf. \appref{sec:derivation} for the full derivation). Over $m$ cycles, the phase $\varphi_+ = \frac{1}{2}(\varphi_1+\varphi_2)$ determines whether successive pairs of transitions interfere constructively. Note that following \eqref{eq:phase_differences}, $\varphi_1$ and $\varphi_2$ have opposite sign. In the nonadiabatic regime $\delta \ll 1$, full population swaps between qubit 1 and 2 occur at the constructive interference condition $\varphi_+ = n\pi$, $n\in\mathbb{Z}$. The frequency of population swaps is set by $\varphi_-=\tfrac{1}{2}(\varphi_2-\varphi_1)$, as depicted in \figref{fig:fig2}d. In contrast to Refs.~\cite{Oliver2005Dec,Sillanpaa2006May,Bjorkman2025Feb}, the effective driven two-level system during the LZ interferometry is not the physical qubit, but consists of the nonlocal bright state and the coupler. This means that a two-qubit gate can be implemented without modulating or driving the data qubits.

\begin{figure*}[ht]
    \centering
    \includegraphics[width=\textwidth]{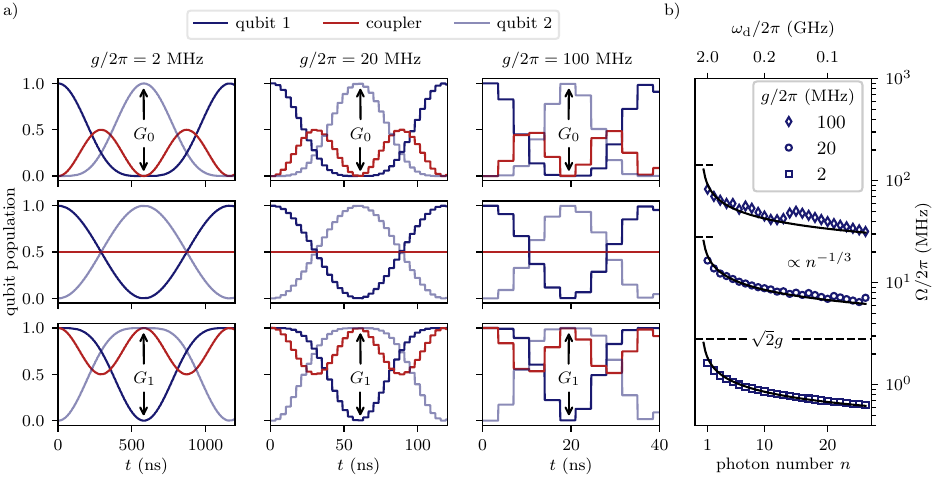}
    \caption{
 \textbf{Simulated population oscillations in the LZ interferometer.} \textbf{a)} Coherent population oscillations from qubit~1 (dark blue) to qubit~2 (light blue) via the coupler (red) are plotted for three different coupling strengths $g$ (columns). The coupler initial state is varied from row to row, resulting in different oscillation patterns, but consistent population swap periods. The parameters used in the simulation are $\omega_\text{q}/2\pi = 5$ GHz, $\epsilon_0/2\pi = 2$ GHz and $n = \epsilon_0/\omega_\text{d} = 7$. The oscillations become more step-like for increasing $g/\omega_\text{d}$. \textbf{b)} The data qubit oscillation frequency $\Omega$ is plotted versus photon number $n$ (bottom x-axis) and the corresponding drive frequency $\omega_\text{d}$ (top x-axis) for the three different coupling strengths depicted in subpanel~\textbf{a}. The population swap frequency $\Omega$ decays algebraically $\propto n^{-1/3}$.}
    \label{fig:fig3}
\end{figure*}

\subsection{Interferometric population oscillations}

The action of the two-qubit gate in \eqref{eq:lzsm_cycle} consists in symmetrically loading and emptying the coupler population from and to the data qubits. To illustrate this, we calculate the dynamics for sinusoidal coupler modulation $\epsilon(t) = \epsilon_\text{d} \sin (\omega_\text{d}t) + \epsilon_0$. We plot the resulting population oscillations in \figref{fig:fig3}{a} versus time for different coupler initial populations at the constructive interference condition $\varphi_+ = 7 \pi$. The Rabi-like oscillations are smooth for $g\ll\omega_\text{d}$, and become step-like for lower drive frequencies, reflecting the discrete nature of the LZ transitions. 

The key property of the three-qubit population oscillations is that for constructive interference, the coupler population always returns to its initial value while at the same time, the data qubit populations invert. For sinusoidal coupler modulation, \eqref{eq:phase_differences} simplifies to
\begin{equation}
\varphi_+ =\frac{1}{2} \int_{t'_{\mathrm{LZ}}}^{t'_{\mathrm{LZ}}+\frac{2\pi}{\omega_\text{d}}} \epsilon (t)\text{d}t 
 = \frac{\epsilon_0}{\omega_\text{d}} \pi, 
 \label{eq:cancellation}
\end{equation}
such that the sum of bath phases $\varphi_1$ and $\varphi_2$ becomes independent of the drive amplitude and the constructive interference condition reads
\begin{equation}
    n = \frac{\epsilon_0}{\omega_\text{d}},
    \label{eq:energy_conservation}
\end{equation}
which can be interpreted as the energy conservation condition for $n$ drive photons matching the energy difference $\epsilon_0$ between the coupler and the qubits. Corrections to the resonance condition \eqref{eq:energy_conservation} arise from nonlinearities or non-sinusoidal modulation of the coupler, as discussed in detail in \appref{sec:nonlinear}. 

We simulate the maximal oscillation frequency $\Omega$ with which the data qubits invert versus the drive frequency $\omega_\text{d}$ for different drive photon numbers $n$ and plot the results in \figref{fig:fig3}b. Starting from a maximum of $\sqrt{2}g$, $\Omega$ decreases algebraically $\propto n^{-1/3}$, consistent with the semi-classical Rabi model. This is in contrast to dispersively mediated multiphoton transitions for which the efficiency of the drive decays exponentially~\cite{Schirk2025Jul, Xia2025Dec}. The difference originates from the fact that the LZ drive is not a perturbation, since $\epsilon_{\mathrm{d}} \gg \Delta$.

We would like to point out several consequences of LZ interference gates which can be inferred directly from the Mach-Zehnder analogy: \\
\textbf{(I)} The data qubits are subjected to LZ interference without being modulated by a flux bias line. It is only the coupler qubit that mediates this modulation, therefore the data qubits can stay idle at their sweet-spots. \\
\textbf{(II)} From \figref{fig:fig3}, it becomes clear that two-qubit gates can be played at arbitrary drive frequencies $\omega_{\mathrm{d}}$ since the constructive interference condition can be fulfilled by tuning the drive amplitude $\epsilon_{\mathrm{d}}$ and offset $\epsilon_{0}$ to match the respective interference condition for a given $\omega_\text{d}$ or vice versa. This opens the possibility for multiplexing flux control pulses by tuning combinations of $\epsilon_{\mathrm{d}}$ and $\epsilon_0$ for different coupler elements to constructive or destructive interference at the same time as detailed in \appref{sec:multiplexing}. \\
\textbf{(III)} The coupler qubit plays an active part in the gate and its degree of freedom needs to be accounted for in the gate analysis. This is in contrast to more conventional gate mechanisms, where leakage to the coupler is generally considered a liability, resulting in coherent errors~\cite{Sung2021Jun}. Here, the coupler can act as a resource providing additional quantum control via preparation and measurement, as discussed in the next section.

\subsection{Two-qubit gates}

When the coupler returns to its initial state at time 
\begin{equation}
    t_\text{swap}=\frac{1}{2\Omega},
\end{equation} $M^m$ effectively mediates a two-qubit gate in the subspace of the data qubits, whose evolution for $\ket{00}, \ket{01}, \ket{10}, \ket{11}$ is found to be of the form 
\begin{equation}
G_0 = 
\begin{pmatrix}
    1 & 0 & 0 & 0 \\
    0 & 0 & -1 & 0  \\
    0 & -1 & 0 & 0  \\
    0 & 0 & 0 & -1
\end{pmatrix} 
\label{eq:G0}
\end{equation} and
\begin{equation}
G_1 = 
\begin{pmatrix}
    -1 & 0 & 0 & 0 \\
    0 & 0 & -1 & 0  \\
    0 & -1 & 0 & 0  \\
    0 & 0 & 0 & 1
\end{pmatrix}, 
\label{eq:G1}
\end{equation}
where the subscript denotes the coupler initial state, which determines the excitation subspace in which the state evolution takes place (cf. \figref{fig:fig1}c). Both gates are locally equivalent (within single qubit $\pi/2$ rotations) to iSWAP, which is a maximally entangling gate. For arbitrary coupler polar angle $\theta$, the only difference is that the phase rotations differ from $\pi$ (cf. \appref{sec:derivation}).

Moreover, playing the gate for a time $t_\text{swap} / 2$ opens possibilities for potentially interesting scenarios as it creates entangled data qubit states either deterministically, or via postselection of the coupler state.  
For example, the initial states
\begin{align}
    \ket{\psi_{1}}_i &= \ket{010}
\end{align}
and
\begin{align}
    \ket{\psi_{2}}_i &= \ket{101}
\end{align}
yield the final states
\begin{align}
    \ket{\psi_{1}}_f &= \frac{ie^{-i\phi}}{\sqrt{2}} \left( \ket{100} + \ket{001}\right) 
    \label{eq:Bell_1}
\end{align}
and
\begin{align}
    \ket{\psi_{2}}_f &= \frac{ie^{i\phi}}{\sqrt{2}}  \left( \ket{110} + \ket{011} \right),
    \label{eq:Bell_2}
\end{align}
with the phase factor $\phi$ depending on the exact drive parameters (see \appref{sec:derivation}). For $\ket{\psi_1}$, the coupler gifts its photon with equal amplitude to the data qubits. For $\ket{\psi_2}$, the coupler receives \textit{half} a photon from each data qubit. Either way, the gate prepares the data qubits in a symmetric Bell state and the coupler can be traced out.

Another example scenario starts from the initial states
\begin{align}
    \ket{\psi_{1}}_i &=  \ket{100}
\end{align}
and
\begin{align}
    \ket{\psi_{2}}_i &=  \ket{110},
\end{align}
giving the final states
\begin{align}
    \ket{\psi_{1}}_f &=  \frac{1}{2}  \left(\ket{100} - \ket{001} \right) + \frac{ie^{i\phi}}{\sqrt{2}} \ket{010}
    \label{eq:entangled}
\end{align}
and
\begin{align}
    \ket{\psi_{2}}_f &= \frac{1}{2}  \left(\ket{110} - \ket{011} \right) + \frac{ie^{-i\phi}}{\sqrt{2}} \ket{101}.
\end{align}
If at the end of the gate the coupler is measured in its initial state $\ket{0}$, the data qubits in $\ket{\psi_1}$ collapse to a singlet Bell state. If, however, the coupler is measured in $\ket{1}$, both data qubits collapse to the product state $\ket{00}$. As a result, the coupler is entangled to whether or not the data qubits are entangled. The same logic applies for $\ket{\psi_2}$. 

Finally, we would like to point out that in all scenarios, because the gate is non-dispersive, control over the coupler degree of freedom either in the form of quantum state preparation or measurement fidelity is essential. Since measurement fidelities are currently lagging behind two-qubit gate fidelities, this can be seen as a downside of the LZ gate mechanism. However, in order to restrict the two-qubit gate to $G_0$, all that is required is a \textit{cold} coupler, which can also be obtained by means other than quantum measurement~\cite{Geerlings2013Mar, Viitanen2024Jun}.

\section{Experimental results}
\label{sec:experiments}

\subsection{Interferometric population transfer}

\begin{figure*}[t]
    \centering
    \includegraphics[width=\textwidth]{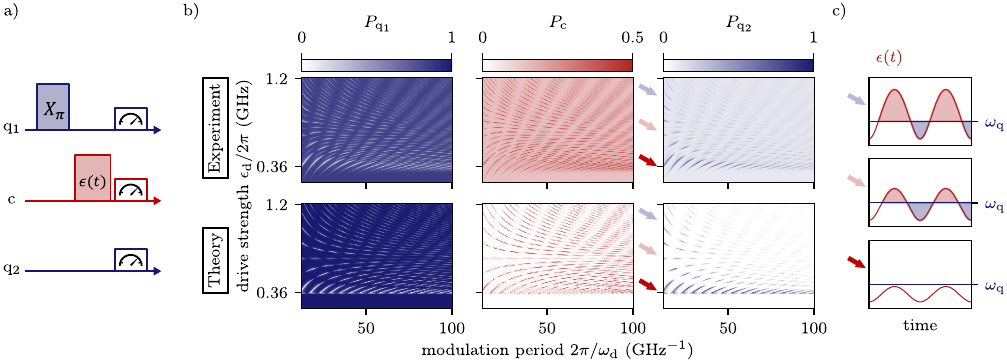}
    \caption{
 \textbf{Three-qubit interferometric population transfer.} \textbf{a)} The pulse sequence for the demonstration of population transfer starts with a $\pi$-pulse on qubit 1 with subsequent coupler flux modulation $\epsilon(t)$ (red). Finally, qubits and coupler are measured simultaneously on separate readout lines. \textbf{b)} The experimental results for a parameter sweep of the inverse drive frequency (modulation period) $2\pi / \omega_{\mathrm{d}}$ (x-axis) and the drive strength $\epsilon_{\mathrm{d}}$ (y-axis) in the first row are compared to the theoretical predictions (cf. \eqref{eq:lzsm_cycle}) in the second row. The data qubit inversion is plotted in blue and the coupler inversion in red. Population is transferred from qubit 1 to the coupler and qubit 2, following a typical interferometric pattern resulting from the interplay of constructive interference (hyperbolic lines) and the modulation of $\Omega$ (diagonal lines). \textbf{c)} Example pulse shapes for three different drive strengths corresponding to the colored arrows in \textbf{b}. Population transfer can be observed once the drive strength exceeds $\epsilon_\text{d} > \epsilon_0$ (red arrow). }
    \label{fig:fig4}
\end{figure*}

We validate the LZ gate mechanism by demonstrating population transfer between the qubits. The data qubits are tuned on resonance to $\omega_{\mathrm{q}} / 2\pi = 3.391\,\mathrm{GHz}$ and the coupler is detuned by $\omega_0 / 2\pi= -359 \,\mathrm{MHz}$, ensuring qubits and coupler are idle at their flux-insensitive sweet-spots. As sketched in \figref{fig:fig4}a, we prepare qubit 1 with a $X_\pi$-pulse and keep the coupler and qubit 2 close to the ground state with thermal populations less than 5 \%. Next, we play a sinusoidal pulse with length $t_{\mathrm{p}} = 300 \, \mathrm{ns}$ on the flux bias line which modulates the coupler spectrum with drive strength $\epsilon_{\mathrm{d}}$ and frequency $\omega_{\mathrm{d}}$, and subsequently read out the qubits and the coupler simultaneously. The measured qubit and coupler inversions are plotted in the first row of \figref{fig:fig4}b. They match the expected interference patterns from the Mach-Zehnder model (\eqref{eq:lzsm_cycle}), which are plotted in the second row for comparison. For the theoretical predictions, we use the fitted flux-tunable coupler spectrum and avoided level crossings to numerically calculate the LZ velocity $v$ and the phases $\varphi_1$ and $\varphi_2$ for each $\epsilon_{\mathrm{d}}$ and $\omega_{\mathrm{d}}$ (cf. \appref{sec:nonlinear}). 

Population swapping is observed for $\epsilon_{\mathrm{d}} \ge \epsilon_0 \approx 2\pi \times 0.36\, \mathrm{GHz}$ (red arrow in \figref{fig:fig4}b \& c), as expected from the LZ mechanism. If $1/\Omega$ is smaller than $t_\text{p}$, full population inversion is observed for qubit 1 and qubit 2 , whereas the coupler is loaded only to $P_\text{c} = 0.5$ because it mediates the population swapping symmetrically and empties with equal rate as it is loaded. Since $\Omega$ decreases with $\epsilon_\text{d}$ and $\omega_\text{d}^{-1}$, the fringe visibility for fixed $t_\text{p}$ becomes fainter towards the top right corners in \figref{fig:fig4}b. Full inversion can be restored in this range by increasing $t_\text{p}$ as long as the qubit decoherence budget allows it. By changing $\epsilon_0$, the entire interferogram can be shifted up or down along the $\epsilon_\text{d}$-axis, providing an additional tuning knob. As we show in \appref{sec:multiplexing}, these control knobs for the interferogram enable either full or zero population swap between the qubits at any drive frequency, opening the way for frequency-multiplexed control of couplers.

\begin{figure}[t]
    \centering
    \includegraphics[width=\linewidth]{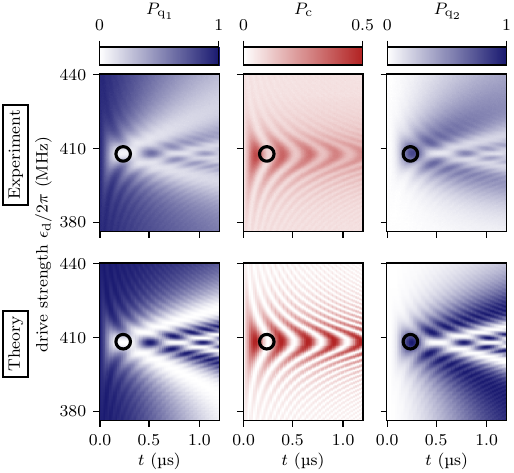}
    \caption{  \textbf{Population oscillations via LZ interference.} Measured population oscillations between the data qubits and the coupler are shown around a constructive interference resonance for $\omega_{\mathrm{d}}/  2\pi = 50\, \mathrm{MHz}$ versus gate time $t$ for the same pulse sequence as in \figref{fig:fig4}a. The data qubit inversions are plotted in blue and the coupler qubit inversion in red. The theory prediction (cf. \eqref{eq:lzsm_cycle}) is shown for comparison in the second row. At resonance, full population inversion for qubit 1 and 2 can be achieved with the coupler returning to its initial state. The black circles mark $t_\text{p} \approx 250$ ns which implements the two-qubit gate $G_0$.}
    \label{fig:fig5}
\end{figure}

To demonstrate the LZ gate in time-domain, we choose a LZ resonance at $\omega_\text{d} = 2\pi \times 50$ MHz and $\epsilon_\text{d}\in2\pi\times  [380, 440]$ MHz and apply the same pulse sequence as in \figref{fig:fig4}a with varying gate time $t$. The measured population oscillations are shown in \figref{fig:fig5} in the first row, and the theoretical predictions in the second row. The resulting oscillations show coherent population transfer from qubit 1 to qubit 2 via the coupler. Because $\epsilon_\text{d}\gg g$, many modulation periods are needed to achieve a population oscillation period, masking the discreteness of the LZ transitions, similar to the regime discussed in the first column in \figref{fig:fig3}a. The time indices highlighted by the black circles at $t_\text{p} \approx 250$ ns in \figref{fig:fig5} correspond to the coupler returning to its initial state and the implementation of the two-qubit gate $G_0$ (cf. \eqref{eq:G0}).

\subsection{Population swap benchmarking}

\begin{figure}[htbp]
    \centering
    \includegraphics[width=\linewidth]{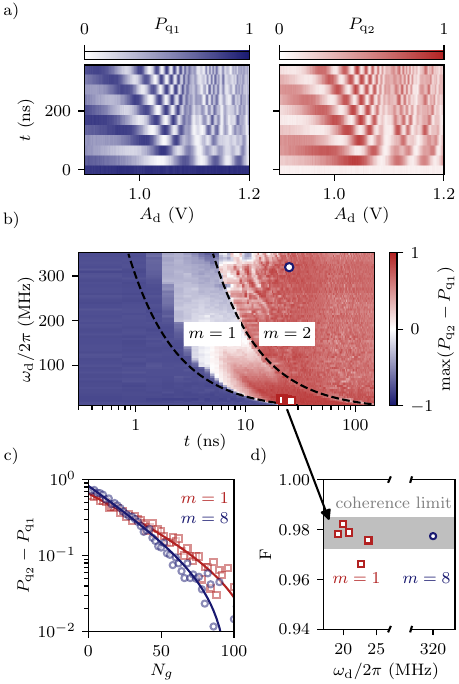}
    \caption{\textbf{Population swap benchmarking on coupled transmons.} \textbf{a)} Population oscillations of qubits 1 and 2 versus room temperature drive amplitude $A_\text{d}$ and time $t$. The drive frequency is $\omega_\text{d} = 2\pi \times 28 \, \text{MHz}$. The swaps correspond to the gate $G_0$. The oscillation frequency increases with increasing amplitude until $A_\text{d}\approx 1.1V$ is reached, where the oscillation frequency goes to zero due to coherent destruction of tunneling~\cite{Grifoni1998Oct}.
    \textbf{b)} The maximal population transfer from qubit 1 to qubit 2 is plotted versus $\omega_\text{d}$ (y-axis) and $t$ (x-axis). Each point corresponds to a calibrated population swap after optimizing $A_\text{d}$. The measurements show that it is possible to calibrate a gate for any drive frequency up to $\omega_\text{d}/2\pi < 350$ MHz when at least $m=2$ LZ cycles are completed (right black dashed line). For lower drive frequencies, i.e. smaller adiabaticity, $m=1$ is sufficient to implement a population swap (left dashed line). The colored markers correspond to the drive parameters for the benchmarking results in \textbf{c} and \textbf{d}.
    \textbf{c)} Population inversion for a gate sequence of an even number $N_g$ of pulses, involving $m=1$ (red) and $m=8$ (blue) drive periods. From exponential fits (solid lines) we extract the swap fidelity versus $\omega_\text{d}$ shown in \textbf{d}.
    \textbf{d)} The measured fidelity is within the range corresponding to an effective decoherence time $T_2^\text{eff} = 2 \pm0.5$ $\upmu$s (gray area).}  
    \label{fig:fig6}
\end{figure}

To benchmark the population swap fidelity corresponding to the two-qubit gate $G_0$ (cf.~Eqs.~\ref{eq:G0}), we implement it on a monolithic device~\cite{QSolid} which uses flux-tunable transmons with similar coherence but increased coupling strength $g/2\pi \approx 300$~MHz. This allows us to circumvent the current coupling limitation of the modular architecture~\cite{Ihssen2025Mar} ($g/2\pi \approx 3$~MHz) and speed up the gate by order of magnitude.

We tune the qubits on resonance at $\omega_\mathrm{q}/2\pi = 5.891$~GHz and calibrate the coupler idle position $\epsilon_0$ to cancel the direct transverse interaction of the data qubits~\cite{qruise_qruiseos}. Similar to \figref{fig:fig4}a, we excite qubit 1 and play a sinusoidal gate pulse with variable duration $t$ and measure the populations of the data qubits simultaneously. Since now the coupler does not have a dedicated readout resonator, its full flux-dependent spectrum is unknown and we refer to the room-temperature drive amplitude $A_\mathrm{d}$ instead of a drive strength $\epsilon_\mathrm{d}$. 

We demonstrate population swaps by tuning $\omega_\mathrm{d}$ and $t$ to observe population inversion. An example swap pattern is shown in \figref{fig:fig6}a. Below $A_\text{d} \approx 1$ V, the coupler does not reach the data qubit resonance, and the resulting population oscillations stem from residual interactions because the coupler approaches the vicinity of the avoided crossings. Beyond $A_\text{d} \approx 1$ V, the pattern shows three characteristic lobes versus $A_\text{d}$ corresponding to the modulation of the population swap frequency resulting from the interferometric LZ mechanism.

To validate the in situ tunability of $\omega_\mathrm{d}$, we perform population swap measurements for various drive frequencies in the interval of $10$\,MHz\,$<\omega_\mathrm{d}/2\pi < 350\,$MHz and plot the maximal data qubit inversion max($P_\mathrm{q2} - P_\mathrm{q1}$) in \figref{fig:fig6}b. The data shows that it is possible to calibrate a drive amplitude for arbitrary $\omega_\mathrm{d}$ to achieve population inversion. Furthermore, the crossover between $m=1$ and $m=2$ drive periods, at which the number of LZ transitions doubles, is visible. The minimal gate duration is set by the time the coupler needs to traverse the first two LZ transitions. For $g > \omega_\text{d}$, full inversion can already be achieved after traversing a single interferometer (cf. \figref{fig:fig2}c), therefore even faster gates (up to the limit shown in \figref{fig:fig3}b) could be achieved by optimizing the pulse shape beyond a sinusoidal waveform.

We calibrate combinations of $A_\text{d}$ and $\omega_\text{d}$ for $m=1$ and $m=8$. We minimize the coherent error contribution by calibrating $A_\text{d}$ and fit the resulting exponential decay curve to $P_\mathrm{q2} - P_\mathrm{q1} = aF^{N_g}+c$ to extract the population swap fidelity $F$. The parameters $a$ and $c$ account for state preparation and measurement errors. We plot two example decay curves, including the corresponding fits, in \figref{fig:fig6}c.

Considering the additional wait time of $16$ ns between the gate pulses, the fitted population swap fidelity $F \approx 0.98$ is limited by our current level of decoherence, $T_2^\text{echo} = 2.2$ $\upmu$s. Assuming qubits and couplers with two orders of magnitude larger coherence, even when considering the impact of asymmetries, decoherence and leakage (see \appref{sec:coherent_errors}, \appref{sec:leakage} and \appref{sec:incoherent_errors}), we expect that the LZ-interferometric gate can reach 99.9 \% population swap fidelity. 

\section{Conclusions}
\label{sec:conclusions}
In this work we have developed a novel parametric two-qubit gate for superconducting qubits. It is based on interference between consecutive Landau-Zener transitions of a flux-tunable coupler through the symmetric quantum superposition of data qubit states. Even though nondispersive in nature, the coupler can be calibrated to return to its initial state at the same time the data qubits invert, rendering the effective two-qubit gate locally equivalent to iSWAP. Furthermore, when the coupler starts in the excited state and the qubits in the ground state, the LZ gate can be used to prepare a Bell state.

We have experimentally demonstrated population swaps on two different platforms, a modular flip-chip architecture of flux qubits separated on different chips, as well as on a monolithic coplanar waveguide transmon architecture. Benefitting from the increased coupling in the monolithic architecture, we find an average population swap fidelity $F \approx 0.98$, limited by decoherence. The key features of the LZ gate are the in situ tunability of its drive frequency, its fast population swap frequency approaching the $\sqrt{2}g$-limit, and the fact that no drive or flux pulse needs to be applied to the data qubits. 
The LZ gate combines the speed of resonant Rabi oscillations with the versatility of parametric gates operating at an arbitrary, tunable drive frequency, thereby offering distinct advantages for scaling superconducting quantum hardware.

\vspace{1mm}

\section*{Acknowledgements}
 We are grateful to Lucas Radtke and Silvia Diewald for technical assistance, and Alessandro Ciani for insightful discussions. 
 This project was funded by the Federal Ministry of Research, Technology, and Space (BMFTR) within project QSolid (FKZ:13N16149 and 13N16151).
 N.Z.~acknowledges funding from the Deutsche Forschungsgemeinschaft (DFG – German Research Foundation) under project number 450396347 (GeHoldeQED).
 Facilities use was supported by the KIT Nanostructure Service Laboratory. 
 We acknowledge the measurement software framework qKit.

\section*{Appendices}

\twocolumngrid
\appendix
\setcounter{equation}{0}
\setcounter{figure}{0}
\setcounter{table}{0}

\makeatletter
\renewcommand{\theequation}{A\arabic{equation}}
\renewcommand{\thefigure}{A\arabic{figure}}
\renewcommand{\theHfigure}{A\arabic{figure}}
\renewcommand{\thetable}{A\arabic{table}}
\renewcommand{\thesection}{\Alph{section}}

\section{Derivation of the gate unitaries}
\label{sec:derivation}

\begin{figure*}[htbp]
    \centering
    \includegraphics[width=1.0\textwidth]{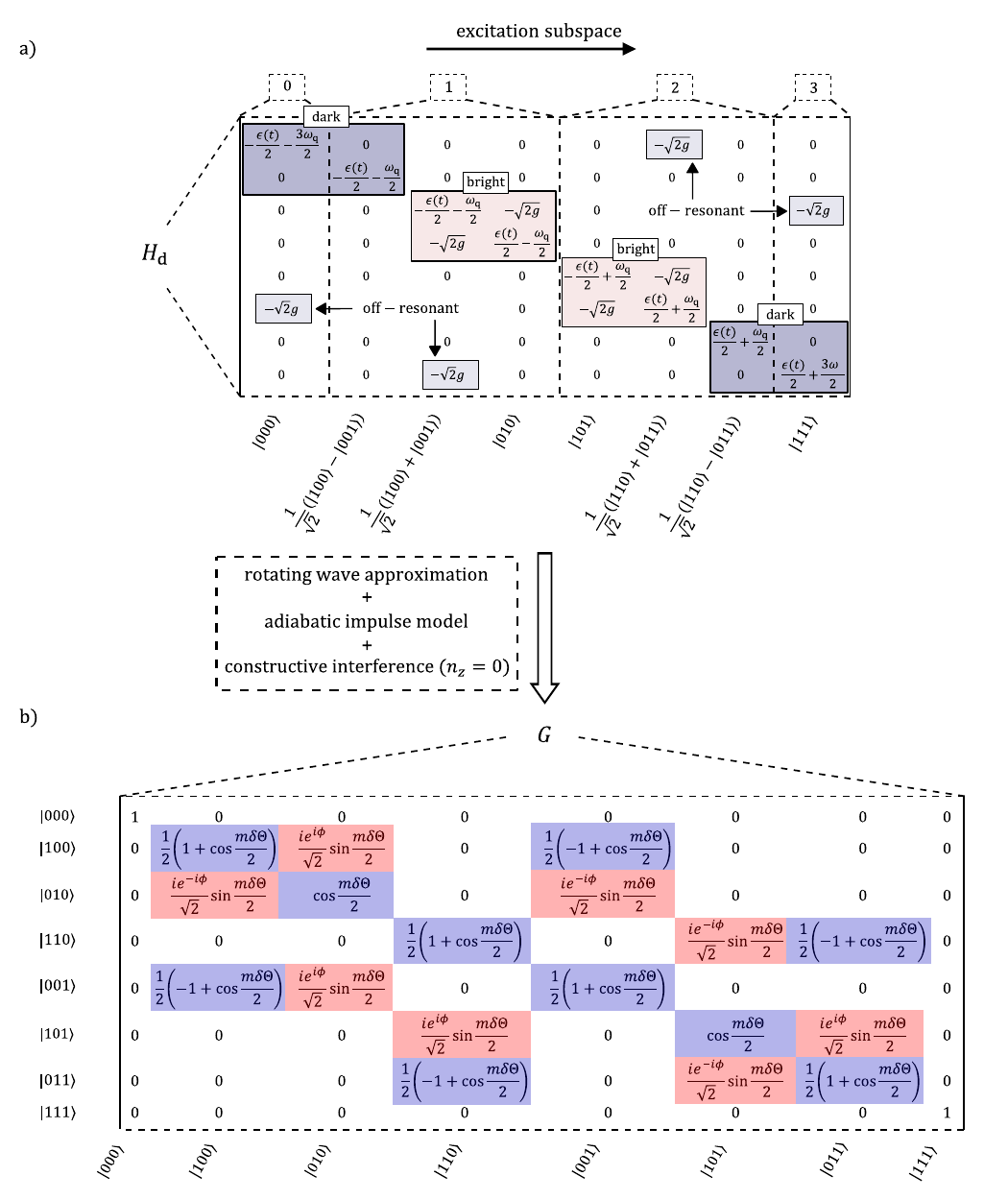}
    \caption{
    \textbf{Hamiltonian in the diabatic basis and evolution operator in the local basis.} 
    \textbf{a)} The diabatic Hamiltonian $H_\text{d}$ shows the coupled blocks revealing the LZ crossings between the bright states (bright red), the decoupled dark states (dark blue) and the off-resonant coupling between different excitation subspaces (bright blue).
    \textbf{b)} The evolution operator for $m$ drive periods shows sinusoidal population oscillations between the data qubits (blue) via the coupler (red). To recover $G_0$ and $G_1$ from the main text, the red terms must vanish.
    }
    \label{appfig:figS1}
\end{figure*}

We transform the local Hamiltonian of main text \eqref{eq:H_local} into the diabatic basis (cf. \figref{fig:fig1}c) via the basis transformation
\begin{equation}
    S = 
    \begin{pmatrix}
        1 & 0 & 0 & 0 & 0 & 0 & 0 & 0 \\
        0 & \frac{1}{\sqrt{2}} & \frac{1}{\sqrt{2}} & 0 & 0 & 0 & 0 & 0 \\
        0 & 0 & 0 & 1 & 0 & 0 & 0 & 0  \\
        0 & 0 & 0 & 0 & 0 & \frac{1}{\sqrt{2}} & \frac{1}{\sqrt{2}} & 0 \\
        0 & -\frac{1}{\sqrt{2}} & \frac{1}{\sqrt{2}} & 0 & 0 & 0 & 0 & 0 \\
        0 & 0 & 0 & 0 & 1 & 0 & 0 & 0 \\
        0 & 0 & 0 & 0 & 0 & \frac{1}{\sqrt{2}} & -\frac{1}{\sqrt{2}} & 0 \\
        0 & 0 & 0 & 0 & 0 & 0 & 0 & 1 \\
    \end{pmatrix}.
\end{equation}
Its block-structure is visualized in \figref{appfig:figS1}a: Neglecting the off-resonant terms $\sqrt{2}g \ll 2\omega_{\mathrm{q}}$ (rotating wave approximation), the matrix is block-diagonal and each of the two blocks forms an avoided crossing with gap $\Delta = 2\sqrt{2}g$. When the time-dependent coupler frequency modulation $\epsilon(t)$ traverses $0$, i.e. the coupler crosses the resonance of the data qubits at frequency $\omega_\text{q}$, each block resembles a typical LZ transition. The evolution operator incorporating the transition probability according to \eqref{eq:LZ_probability} as well as the Stokes phase~\cite{Shevchenko2010Jul} (with $\Gamma$ being the gamma function)
\begin{equation}
    \varphi_{\mathrm{s}} = \frac{\pi}{4} + \delta ( \ln{\delta} - 1 ) + \arg{ \Gamma(1 - i\delta)}
\end{equation}
reads
\begin{equation}
\tilde{N} = 
\begin{pmatrix}
     \mathbb{I}_{2 \times 2} & 0 & \cdots & 0   \\
      0 & N & \ddots & \vdots  \\
      \vdots & \ddots & N & 0  \\
      0 & \cdots  & 0 & \mathbb{I}_{2 \times 2} \\
\end{pmatrix}.
\end{equation}
We introduced the notation $\tilde{N}$ to distinguish the 8-level matrix describing the full three-qubit system from the $2 \times2$ LZ transfer matrix $N$ (cf. \eqref{eq:beam_splitter}) that describes the isolated single blocks within the respective excitation subspace (cf. \figref{appfig:figS1}a).

The coupler modulation
\begin{equation}
\epsilon(t) = \epsilon_0 + f(\epsilon_{\mathrm{d}}, \omega_{\mathrm{d}})
\end{equation} 
is in general a periodic and continuous function $f$ with period $T = \frac{2\pi}{\omega_{\mathrm{d}}}$ and maximal amplitude $\epsilon_{\mathrm{d}}$. If $g \ll |\epsilon_{\mathrm{d}} - \epsilon_0|$, the modulation sweeps approximately linearly during the crossing within the region of width $\Delta$ and the definition of a sweep velocity as in \eqref{eq:lz_velocity} is justified.

Within one period of $\epsilon(t)$, LZ transitions are induced twice (compare \figref{fig:fig2}b). By comparing the energy terms in \figref{appfig:figS1}a, the dynamical phase evolution matrix in between crossing points $\delta t_1 = t_{\mathrm{LZ}}''-t_{\mathrm{LZ}}'$ and $\delta t_2 = T + t_{\mathrm{LZ}}'-t_{\mathrm{LZ}}''$ is found to be

\begin{align*}
\tilde{U}_{j} = & e^{i \varphi_j\mathrm{diag}(-1, -1, -1, 1, -1, 1, 1, 1)} \\
   \times & e^{i \frac{\omega_{\mathrm{q}} \delta t_j}{2} \mathrm{diag}(-3, -1, -1, -1, 1, 1, 1, 3)}.
\end{align*}
We write the total evolution matrix for $m$ drive cycles ($2m$ LZ transitions) by neglecting the exact rise-up and ring-down of the coupler pulse before the first and last LZ transition as

 \begin{align}
    \tilde{M}^{m} &= \left(\tilde{U}_{2} \tilde{N}^T \tilde{U}_{1} \tilde{N} \right)^m \\
    &= \underbrace{e^{i m\pi \frac{\omega_{\mathrm{q}}}{\omega}\mathrm{diag}(-3, -1, -1, -1, 1, 1, 1, 3)}}_{\text{rotating frame of the the resonant qubits}} \\ 
    & \times\underbrace{e^{i m \varphi_{+}\mathrm{diag}(-1, -1, 0, 0, 0,0, 1, 1)}}_{\text{phase accumulation of the dark states}}  \\
    &\times 
\underbrace{
\begin{pmatrix}
     \mathbb{I}_{2 \times 2} & 0 & \cdots & 0   \\
      0 & M^{m} & \ddots & \vdots  \\
      \vdots & \ddots & M^m & 0  \\
      0 & \cdots  & 0 & \mathbb{I}_{2 \times 2} \\
\end{pmatrix}.
}_{\text{nonadiabatic interference between the bright states}}
\label{eq:general_formula}
\end{align}
We have therefore split the time evolution into a product of three distinct terms reflecting the different physical meaning (cf. corresponding labels).

The nonadiabatic transitions and interference between the bright states are described by the blocks $M^m$ as given in \eqref{eq:lzsm_cycle}. To find $M^m$ it is instructive to decompose $M$ into Pauli matrices since then we can write
\begin{equation}
    M^m = \cos \left( \frac{m\Theta}{2} \right) \mathbb{I}-i\sin \left( \frac{m\Theta}{2}\right)\hat{n} \cdot \vec{\sigma}
\end{equation}
with $\hat{n} = (n_x, n_y, n_z)$ being the unit rotation axis with the rotation angle given by
\begin{equation}
\cos \frac{\Theta}{2} = \frac{\mathrm{Tr}[M]}{2} \\
\end{equation}
and the components
\begin{equation}
    n_i = \frac{i}{2\sin \left( \frac{\Theta}{2} \right)}\mathrm{Tr}[\sigma_iM].
\end{equation}
Explicitly, we find
\begin{align}
    \cos{\frac{\Theta}{2}} & = P\cos{\varphi_+} + (1-P)\cos(\varphi_- - 2\varphi_s) \label{eq:rotation_angle} \\
    n_x & = \frac{\sqrt{P(1-P)}}{\sin{\frac{\Theta}{2}}} \left( \sin(\varphi_- - \varphi_s) - \sin(\varphi_+ + \varphi_s) \right) \\
    n_y & = \frac{\sqrt{P(1-P)}}{\sin{\frac{\Theta}{2}}} \left( \cos(\varphi_- - \varphi_s) - \cos(\varphi_+ + \varphi_s) \right) \\
    n_z & = \frac{-1}{\sin{\frac{\Theta}{2}}} \left( P\sin{\varphi_+} + (1-P) \sin(\varphi_- - 2\varphi_s) \right) 
\label{eq:Pauli_decomposition}
\end{align} where we introduced

\begin{equation}
    \begin{aligned}
        \varphi_+ &= \frac{\varphi_1 + \varphi_2}{2} \\
        \mathrm{and} \quad \varphi_- &= \frac{\varphi_2 - \varphi_1}{2}.
    \end{aligned}
\end{equation}
as given in the main text.

The LZ cycles rotate the states between $\ket{010}$ to $\ket{100} + \ket{001}$ as well as $\ket{101}$ to $\ket{110} + \ket{011}$ non-trivially around the axis $\hat{n}$ (cf. \figref{fig:fig2}d). The condition for full data qubit inversion is
\begin{equation}
    n_z = 0 
\end{equation}
which coincides with the resonance condition
\begin{equation}
    \varphi_+ = \pi k, \quad \mathrm{for} \quad 1-P \ll 1 \\.
\end{equation}
For further details see Ref.~\cite{Ashhab2007Jun} and~\cite{Shevchenko2010Jul}. Notice that some signs in our expressions deviate from Ref. \cite{Shevchenko2010Jul} because we work in the diabatic basis and our definitions of the phase differences (\eqref{eq:phase_differences}) imply alternating signs $\text{sgn}(\varphi_1) = -\,\text{sgn}(\varphi_2)$ because $\epsilon(t)$ is defined symmetrically around $\omega_{\mathrm{q}}$.

The evolution in the computational basis is found by transforming back via $\tilde{U}_{{\mathrm{local}}} = S\tilde{M}^m S^T$. We calculate $\tilde{U}_{{\mathrm{local}}}$ numerically by integrating \eqref{eq:phase_differences} and evaluating \eqref{eq:general_formula}.

$\tilde{U}_{{\mathrm{local}}}$ simplifies for the resonance condition $n_z = 0$: Full inversion oscillations are achieved for arbitrary $P$ as long as 
\begin{equation}
    n_z = 0,
    \label{eq:resonance_condition}
\end{equation} even if $\varphi_{+} \ne \pi k$. 
We parametrize
\begin{equation}
\frac{\delta\Theta}{2} = \frac{\Theta}{2} - n\pi,
\end{equation}
which leads to
\begin{equation}
\cos\left(m\frac{\Theta}{2}\right)
=(-1)^{mn}\cos\left(m\frac{\delta\Theta}{2}\right).
\end{equation}
We further describe the equatorial component of the rotation axis by introducing the phase $\phi$ via
\begin{equation}
e^{i\phi}=n_x-in_y.
\end{equation}
The local gate unitary can then be written by factoring out the rotating frame and the global phase as
\begin{equation}
G =
e^{-im\pi\frac{\omega_\text{q}}{\omega}
\left(\sigma_z^{\text{q$_1$}}+\sigma_z^{\text{c}}+\sigma_z^{\text{q$_2$}}\right)}
(-1)^{mn}\tilde{M}^m.
\end{equation}
The resulting gate is shown in \figref{appfig:figS1}b and describes symmetric excitation transfer between the data qubits mediated by the coupler.

Complete data qubit inversion occurs when the coupler returns to its initial state when
\begin{equation}
\sin\left(m\frac{\delta\Theta}{2}\right)=0.
\end{equation}
At these points, the dynamics separate into two conditional two-qubit gates depending on whether the coupler starts in $\ket{0}$ or $\ket{1}$. At the first inversion point, $m\frac{\delta\Theta}{2}=\pi$ which corresponds to the gate time $t_\text{swap}$, the two-qubit gates in the data qubit basis are
\begin{equation}
G_0=
\begin{pmatrix}
1 & 0 & 0 & 0 \\
0 & 0 & -1 & 0 \\
0 & -1 & 0 & 0 \\
0 & 0 & 0 & -1
\end{pmatrix}
\label{eq}
\end{equation}
for the coupler starting in the ground state
and
\begin{equation}
G_1=
\begin{pmatrix}
-1 & 0 & 0 & 0 \\
0 & 0 & -1 & 0 \\
0 & -1 & 0 & 0 \\
0 & 0 & 0 & 1
\end{pmatrix}
\label{eq}
\end{equation}
for the coupler starting in the excited state. Both are locally equivalent to iSWAP as can be verified via single data qubit phase rotations 
\begin{equation}
R_z^{\text{q$_1$}}\left(-\frac{\pi}{2}\right)
\otimes
R_z^{\text{q$_2$}}\left(-\frac{\pi}{2}\right)
G_0
=
\begin{pmatrix}
1 & 0 & 0 & 0 \\
0 & 0 & i & 0 \\
0 & i & 0 & 0 \\
0 & 0 & 0 & 1
\end{pmatrix},
\end{equation}
and
\begin{equation}
-R_z^{\text{q$_1$}}\left(\frac{\pi}{2}\right)
\otimes
R_z^{\text{q$_2$}}\left(\frac{\pi}{2}\right)
G_1
=
\begin{pmatrix}
1 & 0 & 0 & 0 \\
0 & 0 & i & 0 \\
0 & i & 0 & 0 \\
0 & 0 & 0 & 1\
\end{pmatrix}.
\end{equation}

For half-oscillation periods ($m\delta \Theta=\pi$, $t = t_\text{swap}/{2}$), the coupler cannot be trivially traced out and the gate in the full basis reads

\begin{equation}
    G = \begin{pmatrix}
        1 & 0 & 0 & 0 & 0 & 0 & 0 & 0 \\
        0 & \frac{1}{2} & \frac{ie^{i\phi}}{\sqrt{2}} & 0 & -\frac{1}{2} & 0 & 0 & 0 \\
        0 & \frac{ie^{-i\phi}}{\sqrt{2}} & 0 & 0 & \frac{ie^{-i\phi}}{\sqrt{2}} & 0 & 0 & 0 \\
        0 & 0 & 0 & \frac{1}{2} & 0 & \frac{ie^{-i\phi}}{\sqrt{2}} & -\frac{1}{2} & 0 \\
        0 & -\frac{1}{2} & \frac{ie^{i\phi}}{\sqrt{2}} & 0 & \frac{1}{2} & 0 & 0 & 0 \\
        0 & 0 & 0 & \frac{ie^{i\phi}}{\sqrt{2}} & 0 & 0 & \frac{ie^{i\phi}}{\sqrt{2}} & 0 \\
        0 & 0 & 0 & -\frac{1}{2} & 0 & \frac{ie^{-i\phi}}{\sqrt{2}} & \frac{1}{2} & 0 \\
        0 & 0 & 0 & 0 & 0 & 0 & 0 & 1 \\
    \end{pmatrix}.
    \label{eq:half_swap_gate}
\end{equation}

\section{Influence of the nonlinear coupler spectrum}
\label{sec:nonlinear}

\begin{figure*}[htbp]
    \centering
    \includegraphics[width=\textwidth]{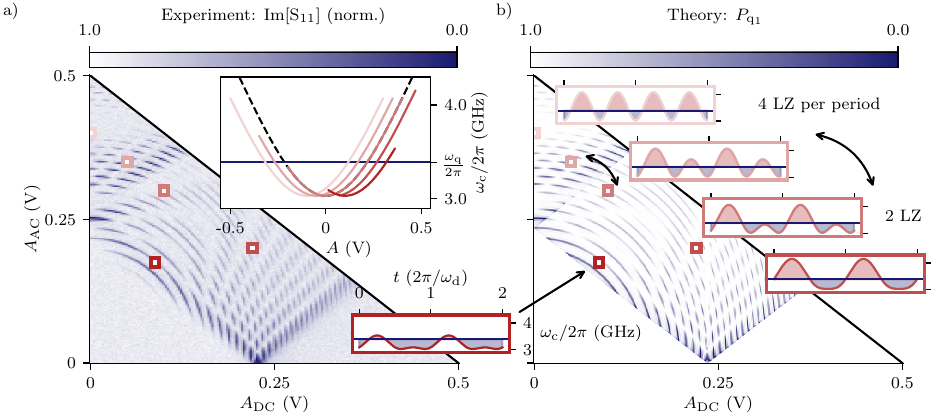}
    \caption[Effect of the nonlinear coupler spectrum on LZ resonances]{\textbf{Effect of the nonlinear coupler spectrum on LZ resonances.} \textbf{a)} Readout response of qubit 1 after a $X_\pi$-pulse followed by a $200$~ns coupler flux pulse with DC amplitude $A_\text{DC}$ and AC amplitude $A_\text{AC}$. Near $\epsilon_0=0$ ($A_\text{DC}\approx0.22$~V), the coupler is modulated within an approximately linear spectral region, whereas increasing $A_\text{AC}$ and approaching the coupler sweet spot ($A_\text{DC}=0$) bends the LZ interference fringes due to spectral nonlinearity. The inset shows the flux-dependent coupler spectrum (black dashed line) versus the total applied voltage $A$. The color coded lines correspond to the spectral regions for the drive parameters indicated by the colored square markers. The lines are shifted on the x-axis for visibility. \textbf{b)} Adiabatic impulse model calculation, in quantitative agreement with the measurement. The inset shows two drive periods for the marked parameters, illustrating the increasingly nonsinusoidal modulation and the resulting modification of the inter-crossing phase accumulation. At large drive amplitudes near the sweet spot, each period traverses four rather than two LZ crossings, effectively doubling the number of LZ cycles. The resulting crossover appears as a diagonal feature in the qubit inversion and is marked by the black double arrow.}
    \label{appfig:figS4}
\end{figure*}

We examine the impact of the nonlinearity of the coupler spectrum by measuring LZ interference similar to the experiment in  \figref{fig:fig4} but for different AC and DC amplitudes of the flux pulse. The experimentally accessible range is limited by $A_\text{DC}+A_\text{AC}=0.5$~V at room temperature. The qubit~1 interferogram in \figref{appfig:figS4}a shows typical LZ interference fringes at the symmetry point $A_\text{DC}\approx0.22$~V, where the coupler modulates symmetrically around $\omega_\text{q}$. The fringes bend towards lower $A_\text{DC}$ as well as for larger $A_\text{AC}$. Eventually they become nearly horizontal. This is due to the increasing nonlinearity of the coupler spectrum, which affects the adiabatic phase accumulation. This changes the constructive interference condition in \eqref{eq:energy_conservation} into an AC-amplitude dependent form, which is responsible for the interference fringes bending horizontally.

To illustrate which parts of the coupler spectrum are sampled, five drive points are marked in \figref{appfig:figS4}a with different colors and the corresponding coupler frequency modulations are shown in the inset. Away from the sweet spot, the nonlinear spectrum distorts the otherwise sinusoidal flux drive into an asymmetric, non-sinusoidal frequency modulation.

As shown in \figref{appfig:figS4}b, the adiabatic impulse model accurately reproduces the resulting fringe bending and deformation. We plot two periods of the calculated coupler modulation versus time in the colored insets, whose color code matches the parameters marked with the colored markers both in \figref{appfig:figS4}a \& b. A qualitative change in the interferogram occurs once the modulation crosses the coupler sweet spot: the frequency susceptibility changes sign within a drive cycle, generating an additional pair of LZ crossings. This transition from two to four LZ events per period appears as a diagonal boundary in the interferogram. It is accompanied by enhanced fringe visibility due to the increased population transfer because of the increase of the number of LZ transitions per time period.

\section{Influence of the coupler degree of freedom}
\label{sec:coupler}

\begin{figure}[htbp]
    \centering
    \includegraphics[width=\linewidth]{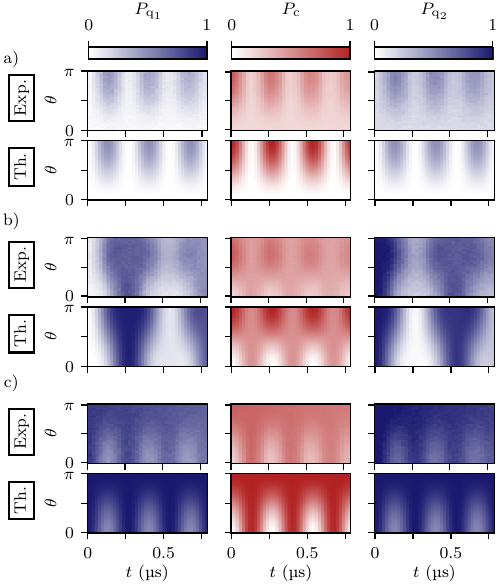}
    \caption{\textbf{Influence of the coupler state on the population oscillations.}
    \textbf{a)} Measured (Exp.) and simulated (Th.) population oscillations for the data qubits in their thermal states close to $\ket{00}$ versus time and coupler state polar angle $\theta$. The excited coupler ($\theta = \pi$) transfers population symmetrically to both data qubits and the oscillations vanish near the coupler ground state because the zero-excitation subspace is decoupled. 
    \textbf{b)} For the data qubits initialized in $\ket{01}$, the excitation swaps between the data qubits through the coupler for all $\theta$, with the oscillation shape depending on the coupler state. 
    \textbf{c)} For the data qubits initialized in $\ket{11}$, the population oscillates symmetrically between the data qubits and the coupler and vanishes near $\theta=\pi$, where the fully excited three-excitation state is decoupled.
    }
    \label{appfig:figS2}
\end{figure}

To probe the dependence of the LZ gate on the initial coupler state, we prepare the coupler at an arbitrary Bloch-sphere polar angle $\theta$ using an $X_\theta$ pulse and initialize the data qubits in different excitation configurations. We then apply the gate pulse shown in \figref{fig:fig5} around $\epsilon_\text{d}/2\pi=410$~MHz, and measure the data qubits as well as the coupler. The measured and simulated populations are shown in \figref{appfig:figS2}.

When one data qubit is initially excited, complete population transfer between the data qubits occurs for any coupler state, only the shape of the intermediate oscillation depends on $\theta$. When both data qubits start in the same state, population is exchanged symmetrically with the coupler and returns to the initial configuration after one oscillation period.

\begin{figure}[htbp]
    \centering
    \includegraphics[width=\linewidth]{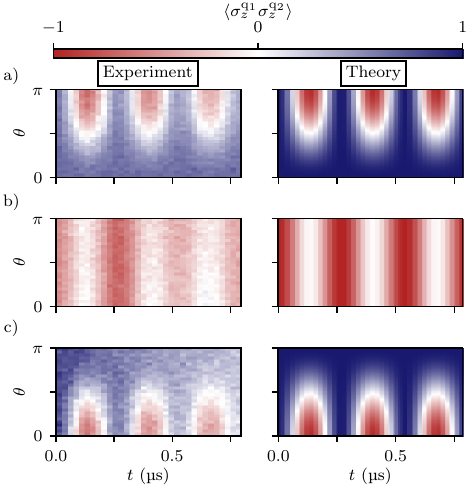}
    \caption{\textbf{Influence of the coupler state on the data qubit correlations.} 
   Qubit-qubit correlations $\langle \sigma_z^{\text{q$_1$}}\sigma_z^{\text{q$_2$}}\rangle$ 
versus coupler pulse duration $t$ and initial coupler polar angle $\theta$, measured (left column) and simulated (right column).
   \textbf{a)} Both data qubits initialized near $\ket{00}$. \textbf{b)} Data qubits initialized in $\ket{01}$ \textbf{c)} Data qubits initialized in $\ket{11}$. Correlation oscillations occur in the one- and two-excitation subspaces but vanish for the zero- and three-excitation states, corresponding to $\theta=0$ in \textbf{a} and $\theta=\pi$ in \textbf{c}. In \textbf{a} and \textbf{c}, the correlation changes sign after half a population oscillation period.
     }
\label{appfig:figS3}
\end{figure}

On top of analyzing individual populations, we consider the correlation $\langle \sigma_z^{\text{q$_1$}}\sigma_z^{\text{q$_2$}}\rangle$ which is positive when equal computational states are favored, negative when opposite states are favored, and vanishes when both outcomes occur with equal probability.

\figref{appfig:figS3} shows the measured and calculated correlations versus the initial coupler polar angle $\theta$ for three data qubit configurations. When only one data qubit is excited, the system evolves simultaneously in the one- and two-excitation subspaces, depending on the coupler state. Correlations therefore oscillate for all $\theta$, but remain negative. 

If the data qubits start in the same state, the behavior changes qualitatively. For $\ket{00}$, the dynamics interpolate between the one-excitation subspace and the dark state $\ket{000}$, causing the oscillations to vanish as $\theta\rightarrow0$. Conversely, for $\ket{11}$ they vanish as $\theta\rightarrow\pi$, where the system approaches the dark state $\ket{111}$. During the evolution, the correlations can change sign and become anticorrelated at half a swap period.

Perfect anticorrelation alone does not demonstrate entanglement. There exists a separable mixture that produces the same $\sigma_z$-correlations as the Bell states while containing no entanglement. However, since the initial states are close to pure states, i.e. the thermal population is $p_\text{th} <0.05$, \figref{appfig:figS3} indicates that the observed anticorrelations are consistent with the data qubits being in a Bell state according to the main text \eqref{eq:Bell_1} and \eqref{eq:Bell_2}.

\section{State preparation via coupler measurements}
\label{sec:post-selection}

\begin{figure}[htbp]
    \centering
    \includegraphics[width=\linewidth]{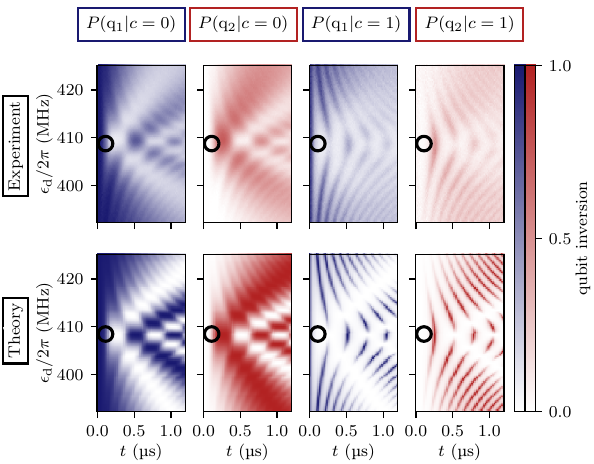}
    \caption{\textbf{Two-qubit state preparation via postselection of the coupler state.} The measurement outcomes for the qubit 1 (blue) and qubit 2 (red) inversion after a pulse sequence similar to \figref{fig:fig4}a are postselected on the final coupler state: $\ket{0}$ in the left two columns, $\ket{1}$ in the right two columns. The measurements are shown in the first row. The theoretical calculations are shown in the second row. The gate for half a swap period is highlighted with black circles.}
    \label{appfig:figS_postselection}
\end{figure}

We demonstrate the preparation of two-qubit states via the post-selection of the coupler state. In the first row of \figref{appfig:figS_postselection} we show post-selected population oscillations corresponding to the pulse sequence shown in \figref{fig:fig4}a. We separate the data qubit measurement outcomes depending on the single shot measurement outcome of the coupler after the pulse sequence. The qubit populations post-selected on whether the coupler is measured in $\ket{0}$ or $\ket{1}$ are $P(\text{q$_{1,2}$}|\text{c}=0)$ and $P(\text{q$_{1,2}$}|\text{c}=1)$. The theoretical calculations taking into account a thermal coupler population of $p_\text{th} = 0.03$ are shown in \figref{appfig:figS_postselection} in the second row. The population oscillations are suppressed when the coupler is measured in $\ket{1}$, opposite to its initial state $\ket{0}$. After half a swap period, both data qubits exhibit $P(\text{q}_1|\text{c}=0)\approx P(\text{q}_2|\text{c}=0) \approx 0.5$, indicating a Bell-type state. Postselection of the coupler on $\ket{1}$ yields $P(\text{q}_1|\text{c}=1)\approx P(\text{q}_2|\text{c}=1) \approx 0$, indicating a product state. This corresponds to the gate shown in \eqref{eq:half_swap_gate} and the final state given in \eqref{eq:entangled} in the main text.

\section{Floquet sidebands}

\begin{figure*}[htbp]
    \centering
    \includegraphics[width=\textwidth]{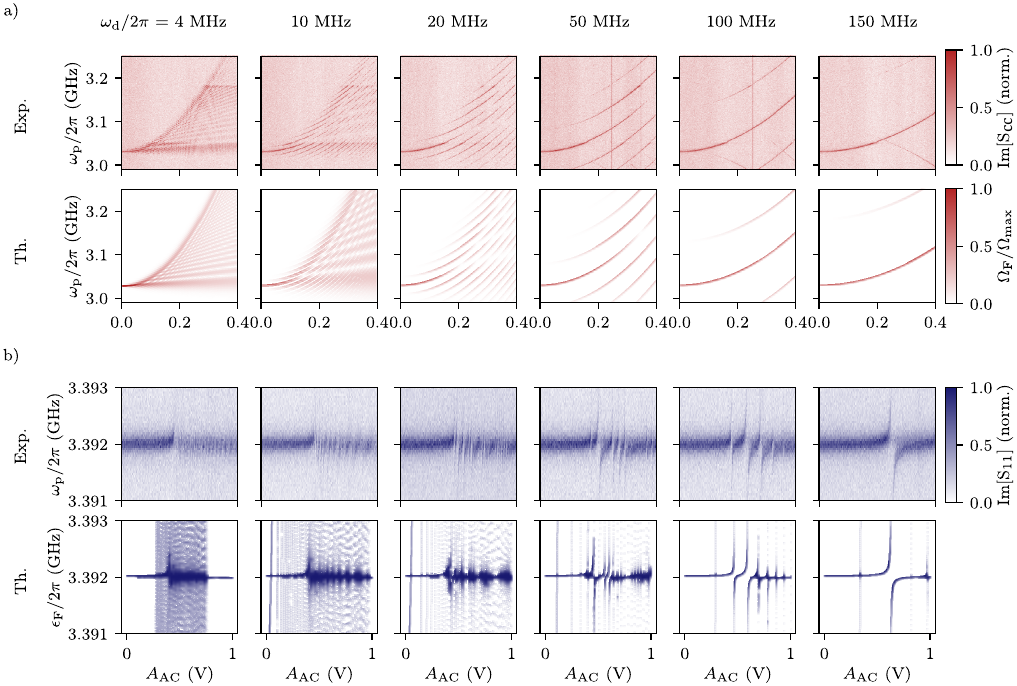}
    \caption{\textbf{LZ resonances as Floquet sidebands.} 
    \textbf{a)} Coupler spectroscopy under flux modulation. The top row shows the measured readout response versus probe frequency $\omega_\mathrm{p}$ and modulation amplitude $A_\text{AC}$ for different drive frequencies $\omega_\text{d}$. The bottom row shows the corresponding normalized Floquet transition matrix elements $\Omega_\text{F}$. Experiment and theory reveal sidebands spaced by $\omega_\text{d}$, Bessel-like amplitude modulation, and a quadratic drive-induced frequency shift. \textbf{b)} Qubit~1 spectroscopy under the same modulation. Avoided crossings appear when coupler Floquet sidebands intersect the data qubit frequency $\omega_\text{q}$, with increasing spacing for larger $\omega_\text{d}$. The bottom row shows the Floquet prediction including transverse coupling $\Omega_\text{F}$ and $N_\text{F}=100$ modes, reproducing the observed crossing positions and widths. For the smallest drive frequencies, the finite Floquet truncation does not capture all features.
    }
    \label{appfig:figS5}
\end{figure*}

A quantum system that is driven periodically can be treated using Floquet theory~\cite{Grifoni1998Oct}, which predicts the emergence of sidebands separated by the drive frequency $\omega_\text{d}$. In the following, we show that the LZ resonances can also be understood as coupler sidebands tuned into resonance with the data qubits. 

To verify the existence of Floquet sidebands, we detune qubit~2 by several GHz from qubit~1 and we probe the qubit~1 and coupler subsystem. We measure two-tone spectroscopy on the coupler while applying a flux drive with amplitude $A_\text{AC}$ and frequency $\omega_\text{FBL}$. The measured spectra are shown in the top row of \figref{appfig:figS5}a.

At $A_\text{AC}=0$, the coupler remains near $3.033$~GHz at its sweet-spot. Increasing $A_\text{AC}$ produces a quadratic frequency shift and additional sidebands whose spacing grows with the drive frequency. Their visibility also oscillates periodically with drive amplitude. These features follow from the approximately parabolic coupler spectrum near the half-flux sweet spot. A sinusoidal flux modulation therefore produces
\begin{equation}
\epsilon(t) \approx
\left[\eta(\omega_\text{FBL})A_\text{AC}
\sin(\omega_\text{FBL}t)\right]^2+\epsilon_0,
\label{eq}
\end{equation}
where $\eta(\omega_\text{FBL})$ accounts for the frequency-dependent attenuation between the applied voltage and the on-chip flux amplitude. This attenuation arises from the finite bandwidth of the flux line and other microwave components and is well described here by
\begin{equation}
\eta(\omega_\text{FBL}) \propto
\frac{1}{\omega_\text{FBL}^2+\tau^2},
\label{eq}
\end{equation}
with a fitted response time $\tau\approx1$~ns.

Expanding the quadratic modulation gives
\begin{equation}
\epsilon(t)=
\frac{\eta^2A_\text{AC}^2}{2}
-\frac{\eta^2A_\text{AC}^2}{2}
\cos(2\omega_\text{FBL}t)
+\epsilon_0.
\label{eq}
\end{equation}
Thus, a flux drive at $\omega_\text{FBL}$ generates a coupler-frequency modulation at $\omega_\text{d}=2\omega_\text{FBL}$ together with a quadratic light shift. These terms explain the sideband spacing and curvature in \figref{appfig:figS5}.

Within the full Floquet description, the corresponding Fourier components are
\begin{align}
F_0 &= \frac{\eta^2A_\text{AC}^2}{2},\\
F_1 &= \frac{\eta^2A_\text{AC}^2}{4},
\end{align}
where $F_0$ shifts the quasienergies and $F_1$ couples neighboring Floquet sectors. The sideband visibility is governed by the transverse $\sigma_x$ transition matrix element, which describes the absorption or emission of multiple drive photons. The calculated matrix elements for $N_\text{Floquet}=100$ modes shown in the bottom row of \figref{appfig:figS5}a reproduce the measured sideband pattern, including the coupling minima associated with coherent destruction of tunneling~\cite{Grifoni1998Oct}.

Because qubit~1 couples transversely to the coupler, each Floquet sideband can hybridize with the qubit when tuned into resonance. We observe this by measuring two-tone spectroscopy on qubit 1 while increasing $A_\text{AC}$ until the sidebands cross $\omega_\text{q}$. The measurements are shown in the top row of \figref{appfig:figS5}b. The avoided crossings are accurately reproduced by the Floquet calculation in the bottom row. Their varying widths reflect the drive-dependent transition matrix elements of the corresponding sidebands.

\section{Coherent errors}
\label{sec:coherent_errors}

\begin{figure}[htbp]
    \centering
    \includegraphics[width=\linewidth]{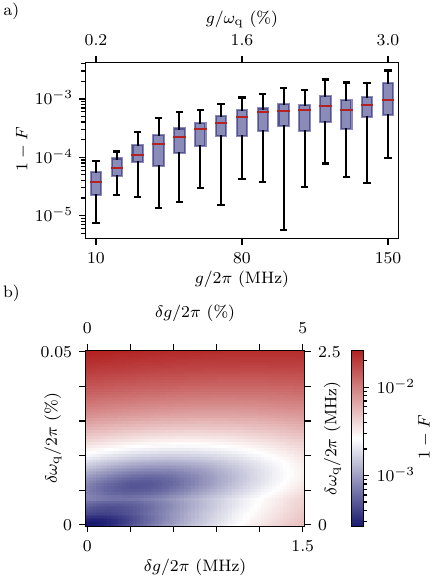}
    \caption{\textbf{Simulated coherent errors.} 
    \textbf{a)} Population swap infidelity $1-F$ versus coupling strength $g$ for calibrations with $n \in \{1,5,10,15,20\}$. Red lines denote the mean, blue bars the standard deviation, and caps the extrema. Infidelity increases at stronger coupling due to additional dynamical oscillations due to the micromotion inbetween LZ transitions.
b) Simulated infidelity for the $n=10$ gate as a function of qubit frequency asymmetry $\delta\omega_\text{q}$ and coupling asymmetry $\delta g$. The symmetric configuration minimizes infidelity, with stronger sensitivity to $\delta\omega_\text{q}$ than to $\delta g$.
} \label{fig:figS6}
\end{figure}

Even for perfectly coherent dynamics, gate fidelities can be limited by intrinsic effects and experimental imperfections. We treat the three dominant coherent error sources: (1) The breakdown of the rotating wave approximation, leading to leakage within the computational subspace, (2) asymmetries in the coupling between the data qubits and the coupler, and (3) frequency asymmetries between the data qubits. We quantify errors stemming from these three contributions by introducing the coupling asymmetry $\delta g=g_{1\text{c}}-g_{2\text{c}}$ and frequency detuning $\delta\omega_\text{q}=\omega_{\text{q}1}-\omega_{\text{q}2}$ and numerically solving the Schrödinger equation for the Hamiltonian in \eqref{eq:H_local}, with qubit~1 initially excited. We fix $\omega_\text{q}/2\pi=5$~GHz and $\epsilon_0/2\pi=2$~GHz and emulate gate calibrations for drive frequencies $\omega=\epsilon/n$, with $n\in \{1,5,10,15,20\}$. For each $n$ and coupling strength $g/2\pi\in(10,150)$~MHz, we sweep the drive amplitude over several resonances, $\epsilon_0<\epsilon_\text{d}<2.5\epsilon_0$, and vary the pulse duration over at least three oscillation periods. The swap fidelity is calculated via
$F=\max(P_{\text{q}2}-P_{\text{q}1})$ and plotted in \figref{fig:figS6}. 

We first simulate the symmetric situation $\delta g = \delta \omega_\text{q} = 0$: \figref{fig:figS6}a shows the mean, standard deviation, and extrema of the resulting infidelity $1-F$. The mean infidelity increases with $g$ but remains below $10^{-3}$ even at $g/2\pi=150$~MHz, corresponding to sub-10-ns two-qubit gates. The large spread between best and worst calibrations indicates that the dominant error is not intrinsic leakage but imperfect sampling of the optimal operating point. Indeed, the minimum infidelity remains below $10^{-4}$ throughout. The simulated values should therefore be regarded as conservative: finer amplitude resolution, an additional frequency correction, or more efficient optimization algorithms could substantially improve calibration.

We further sweep $\delta g$ and $\delta\omega_\text{q}$ in the Hamiltonian and simulate the gate at $g/2\pi=30$~MHz. As shown in \figref{fig:figS6}b, the symmetric configuration yields an infidelity slightly above $10^{-4}$, which increases with either asymmetry. A second local minimum appears where coupling and frequency asymmetries partially compensate each other. The gate is particularly sensitive to frequency mismatch due to its interferometric nature. Flux crosstalk between the flux-bias line and the data qubits will shift the data qubit frequencies in time, leading to a dynamical frequency asymmetry. If this crosstalk is asymmetric, it will contribute in the form of a coherent error as a contribution to $\delta \omega$.

\section{Leakage}
\label{sec:leakage}

\begin{figure*}[htbp]
    \centering
    \includegraphics[width=\textwidth]{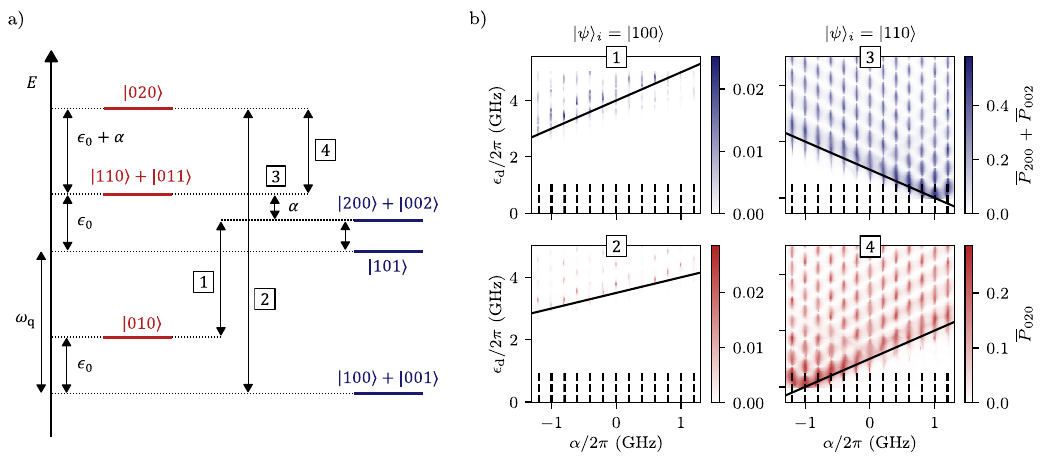}
    \caption{\textbf{Simulated leakage.} 
    \textbf{a)} Bright-state spectrum including the two lowest non-computational levels. Modulated (red) and stationary (blue) levels are separated by the detuning $\epsilon_0$ and anharmonicity $\alpha$. The four dominant leakage crossings are marked by black squares. \textbf{b)} Simulated time-averaged populations $\overline{P}_{200}+\overline{P}_{002}$ (blue) and $\overline{P}_{020}$ (red) versus $\alpha$ and drive amplitude for $\epsilon_0=2$~GHz and $\omega=\epsilon_0/5$. The left and right columns correspond to initial states with one and two excitations, respectively. LZ interference fringes emerge along the four leakage channels identified in \textbf{a}. Solid black lines indicate the corresponding LZ threshold amplitudes, while vertical dashed lines mark the multiphoton resonances.}
        \label{fig:figS7}
\end{figure*}

To identify the impact of leakage out of the computational subspace, we model the system as three coupled qutrits. Following \eqref{eq:H_local}, we use
\begin{equation}
\begin{aligned}
    \frac{H_\text{local}(t)}{\hbar} = \sum_{i = \text{q$_1$,c,q$_2$}} & \left(\omega_\text{q}\ket{1}\bra{1}_i + \left(2\omega_\text{q} + \alpha \right) \ket{2}\bra{2}_i \right) \\
     + \epsilon(t) & \ket{1}\bra{1}_\text{c} +  2\epsilon(t)  \ket{2}\bra{2}_\text{c} \\
     + g\sum_{i = \text{q$_1$,q$_2$}} & ( \ket{0}_\text{c} \bra{1}_i + \beta_{12}\ket{1}_\text{c}\bra{2}_i \\
     &+\beta_{02}\ket{0}_\text{c}\bra{2}_i + \text{h.c.}),
     \label{eq:qutrit_array}
\end{aligned}
\end{equation}
where all elements have the same anharmonicity $\alpha$, while $\beta_{12}$ and $\beta_{02}$ parameterize transitions involving the second excited state. This effective model neglects the flux-dependence of the matrix elements and assumes that the second excited state shifts as $2\epsilon(t)$.

We plot the level structure of the coupled qutrit system in \figref{fig:figS7}a. It reveals four dominant leakage channels:
\begin{enumerate}
\item $\ket{010} \longleftrightarrow \ket{200} + \ket{002}$,
\item $\ket{020} \longleftrightarrow \ket{100} + \ket{001}$,
\item $\ket{110} + \ket{011} \longleftrightarrow \ket{200} + \ket{002}$,
\item $\ket{020} \longleftrightarrow \ket{110} + \ket{011}$.
\end{enumerate}
The first two depend on $\beta_{02}$ and vanish as $\beta_{02}\rightarrow0$. The latter two remain finite because they involve the same single-excitation coupling required for gate operation, but they affect only the two-excitation subspace. The corresponding thresholds for LZ crossings are
\begin{enumerate}
\item $\epsilon_\text{d} > \omega_\text{q} - \epsilon_0 + \alpha$,
\item $\epsilon_\text{d} > \frac{\omega_\text{q}}{2} + \epsilon_0 + \frac{\alpha}{2}$,
\item $\epsilon_\text{d} > \epsilon_0 - \alpha$,
\item $\epsilon_\text{d} > \epsilon_0 + \alpha$.
\end{enumerate}

For capacitively coupled transmons, $\beta_{12}\approx\sqrt{2}$ and symmetry suppresses the direct $0\leftrightarrow2$ transition, giving $\beta_{02}\approx0$. More generally, $\beta_{12}$ is of order unity whereas $\beta_{02}$ is typically small, particularly near symmetry points. During parametric modulation, however, $\beta_{02}$ need not vanish exactly. We therefore simulate $\beta_{12}=\sqrt{2}$ and $\beta_{02}=0.1$, with $\omega_\text{q}/2\pi=5$~GHz, $\epsilon_0/2\pi=1$~GHz, and $\omega=\epsilon_0/5$, while varying $\alpha$.

\figref{fig:figS7}b shows the time-averaged population of $\ket{200}+\ket{002}$ and $\ket{020}$ versus $\alpha$ and drive amplitude. The resulting interference fringes closely follow the predicted resonances and threshold conditions, confirming that these four channels dominate the leakage dynamics.

Leakage from the one-excitation subspace is comparatively weak because channels 1 and 2 are suppressed by the small $\beta_{02}$ and require relatively large drive amplitudes. Their thresholds can therefore be placed above the amplitudes used for computational LZ gates by an appropriate choice of $\omega_\text{q}$, $\epsilon_0$, and $\alpha$. Unlike resonant Rabi gates, larger $|\alpha|$ is not universally beneficial: for negative $\alpha$, increasing $|\alpha|$ can bring $\ket{020}$ closer to $\ket{100}+\ket{001}$ and enhance leakage.

Leakage from the two-excitation subspace is more restrictive. Channels 3 and 4 have thresholds $\epsilon_0-\alpha$ and $\epsilon_0+\alpha$, so changing $\alpha$ suppresses one process while bringing the other closer to resonance. At $\alpha=0$, both coincide with the characteristic computational drive scale $\epsilon_0$. A possible strategy is therefore to choose $|\alpha|>2\epsilon_0$, pushing both leakage channels away from the relevant drive amplitudes. This condition is accessible for strongly anharmonic qubits such as fluxoniums but can be difficult for transmons.

The presence of a leakage resonance does not necessarily imply substantial leakage. As it is the case for LZ interference within the computational subspace, the swapping rate can be calibrated to be zero, enabling coherent suppression of otherwise resonant transitions. Leakage mitigation can therefore be incorporated directly into gate calibration, naturally turning the problem into one of optimal control.

\section{Incoherent errors}
\label{sec:incoherent_errors}

\begin{figure}[htbp]
    \centering
    \includegraphics[width=\linewidth]{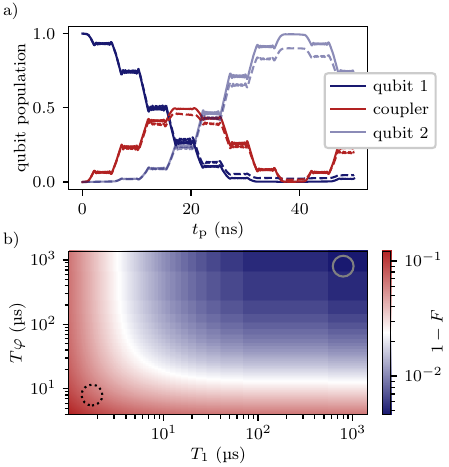}
    \caption{\textbf{Simulated population swap infidelity due to coupler decoherence.} 
    \textbf{a)} Simulated populations of the data qubits (blue) and coupler (red) for short (dashed) and long (solid) coupler coherence times, corresponding to the markers in \textbf{b}. \textbf{b)} Simulated infidelity versus coupler coherence times $T_1$ and $T_\varphi$ for data qubit coherence times of $100~\upmu$s.}
    \label{fig:figS8}
\end{figure}

For the LZ gate, decoherence of the coupler is relevant because it mediates the excitation transfer and contributes to the accumulated LZ phase. We model incoherent errors using the Markovian Lindblad master equation
\begin{equation}
\begin{aligned}
    \dot{\rho}(t) & = -\frac{i}{\hbar} [H_\text{local}(t), \rho(t)] \\
    &+ \frac{\gamma_1}{2}\sum_{i=\text{q$_1$,c,q$_2$}} \left( 2\sigma^i_{-}\rho \sigma^{i}_{+} - \rho \sigma^{i}_{+}\sigma^{i}_{-} - \sigma^{i}_{+}\sigma^{i}_{-} \rho \right) \\
    &+ \frac{\gamma_\varphi}{2}\sum_{i=\text{q$_1$,c,q$_2$}} \left( 2\sigma^i_{z}\rho \sigma^{i}_{z} - 2\rho \right)
    \label{eq:non-unitary}
\end{aligned}
\end{equation}
with the gate Hamiltonian in \eqref{eq:H_local} and relaxation and dephasing channels for all three qubits.

We set the data  qubit coherence times to $T_{1,\text{q}}=T_{\varphi,\text{q}}=100~\upmu$s and vary the coupler $T_1=\gamma_1^{-1}$ and $T_\varphi=\gamma_\varphi^{-1}$ from a few to several hundred microseconds. The parameters are $\omega_\text{q}/2\pi=5$~GHz, $\epsilon_0/2\pi=2$~GHz, $g/2\pi=30$~MHz, and $n=\epsilon_0/\omega=10$. Starting with one excitation in qubit~1, we simulate the resulting population dynamics. 

As shown in \figref{fig:figS8}, short coupler coherence times suppress the swap oscillations and reduce the gate fidelity. Once the coupler coherence exceeds that of the data qubits, however, further improvement is negligible. The gate is therefore limited by the least coherent component of the three-qubit system.

We further find that the simulated fidelity is well described by the heuristic expression
\begin{equation}
F = e^{-t/T_\text{eff}},
\end{equation}
with
\begin{equation}
T_\text{eff} =
\left(
\frac{\gamma_1}{4}
+ \gamma_\varphi
+ \gamma_{1,\text{q}}
+ 2\gamma_{\varphi,\text{q}}
\right)^{-1}.
\end{equation}

Thus, coupler decoherence contributes comparably to data qubit decoherence. High-fidelity operation therefore requires coupler coherence on the same scale as that of the data qubits, whereas substantially exceeding it provides little additional benefit.

\section{Multiplexing}
\label{sec:multiplexing}

\begin{figure}[htbp]
    \centering
    \includegraphics[width=\linewidth]{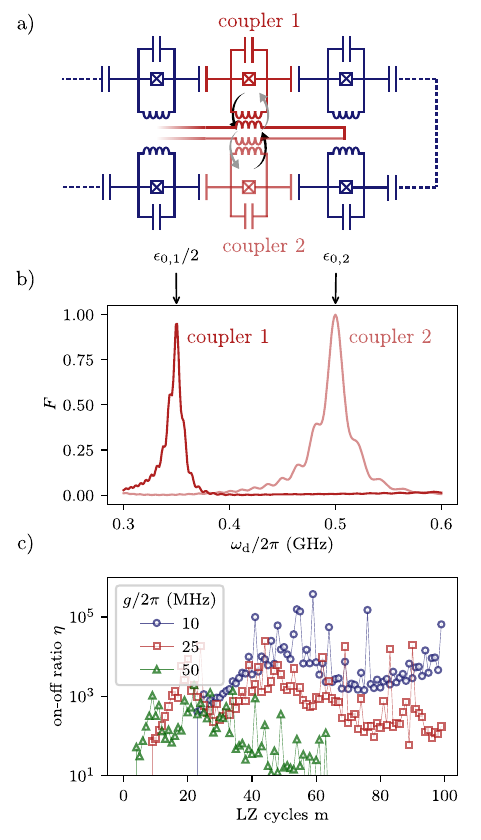}
    \caption{\textbf{Simulated on-off ratio of multiplexed couplers.} 
    \textbf{a)} Circuit diagram of two couplers biased via the same flux bias line. Coupler 1 is detuned by $\epsilon_{0,1}/2\pi = 0.7$ GHz to its adjacent data qubits. Coupler 2 is detuned by $\epsilon_{0,1}/2\pi = 0.5$ GHz. \textbf{b)} Simulated population swap fidelity for a gate containing $m = 30$ LZ cycles played at frequency $\omega_\text{d}$. Coupler 1 (red) implements a population swap when driven at $\epsilon_{0,1}/2$ but not at $\epsilon_{0,2}$, when coupler 2 (light red) implements a population swap. \textbf{c)} Simulated relative on-off ratios between the two couplers versus pulse length in units of LZ cycles $m$ for different coupling strengths $g$.}
    \label{fig:figS9}
\end{figure}

To demonstrate the ability to use constructive and destructive LZ interference to operate multiple couplers with a single flux bias line, we simulate \eqref{eq:H_local} with sinusoidal $\epsilon(t)$ for two couplers experiencing the same coupler flux drive amplitude $\epsilon_\text{d}$ and frequency $\omega_\text{d}$. We choose a qubit frequency of $\omega_\text{q}/2\pi = 5$ GHz and coupler detunings $\epsilon_{0,1}/2\pi = 0.7$ GHz and $\epsilon_{0,2}/2\pi = 0.5$ GHz for coupler 1 and coupler 2, respectively. The circuit is sketched in \figref{fig:figS9}a.

With this drive parameters, constructive interference at coupler 1 can be achieved by using the $n=2$ photon transition by driving at $\omega_{\text{d},1} = \epsilon_{0,1}/2 = 2\pi \times 0.35$ GHz, which is detuned from any constructive interference condition for coupler 2. Vice versa, coupler 2 can be controlled with the $n=1$ photon transition at $\omega_{\text{d},2} = \epsilon_{0,2} = 2\pi \times 0.5$ GHz which is in between the $n=1$ and $n=2$ transition for coupler 1 and therefore destructively interferes at coupler 1.

To quantify the multiplexing performance, we simulate the population transfer $ P_\mathrm{q2} - P_\mathrm{q1}$ from data qubit 1 to data qubit 2 with qubit 1 initially excited for various pulse lengths resulting in $m \in (1, 100)$ LZ periods and drive amplitudes $\epsilon_\text{d} \in (0.5, 2.8)$ GHz ensuring that the LZ transition threshold is reached for both couplers. We define the population swap amplitude for coupler $i$ driven at frequency $\omega_{\text{d},j}$ as $F_{ij}$, where $F$ is defined as $F =  \frac{1}{2}(P_\mathrm{q2} - P_\mathrm{q1}+1)$ to yield $F=1$ for complete population transfer and $F=0$ for no population transfer. We then select the drive amplitudes that yield a population swap gate at the target qubit $i$ by ensuring $F_{ii} > 0.99$. We define the on-off ratio as the ratio of the intended two-qubit swapping induced by coupler $i$ versus the crosstalk-induced population swapping originating from coupler $j$ as $F_{ii} / F_{ij}$ and average over both directions as
\begin{equation}
    \eta = \frac{1}{2} \left( \frac{F_{11}}{F_{12}} + \frac{F_{22}}{F_{21}} \right).
\end{equation}

We plot $F$ versus $\omega_\text{d}$ for $m=30$ and $g/2\pi = 10$ MHz in \figref{fig:figS9}b to illustrate the spectral separation of the LZ gate drives in this multiplexing scheme. The simulated on-off ratio $\eta$ is plotted in \figref{fig:figS9}c versus $m$ for three different coupling strengths $g/2\pi = 10, 25, 50$ MHz. The results show that it is possible to define LZ population swap gates for two couplers simultaneously with an on-off ratio of $\eta>10^3$. Similar to the results in \figref{fig:figS6}a, the simulation includes calibration errors. The on-off ratios therefore decrease with increasing coupling strength $g$, and exhibit oscillatory jumps versus $m$ since optimal parameter points can be missed by the simulation due to the constant parameter sweep step size for $\epsilon_\text{d}$. 

The multiplexing mechanism can be extended to $k$ couplers by finding $k$ parameter pairs $(\epsilon_\text{d}, \omega_\text{d})_k$ that minimize $F_{ij}$ for $j\neq i$ for all $i \in [1, k]$ while at the same time reaching the target fidelities $F_{ii}$.

\bibliography{v1__aipsamp}

\end{document}